# Onshore, offshore or in-turbine electrolysis? Techno-economic overview of alternative integration designs for green hydrogen production into Offshore Wind Power Hubs

Alessandro Singlitico[a], Jacob Østergaard[a], Spyros Chatzivasileiadis[a]

[a] Center for Electric Power and Energy (CEE), Department of Electrical Engineering, Technical University of Denmark (DTU), 2800 Kgs. Lyngby, Denmark.

Corresponding author: alesi@elektro.dtu.dk

**Abstract**

Massive investments in offshore wind power generate significant challenges on how this electricity will be integrated into the incumbent energy systems. In this context, green hydrogen produced by offshore wind emerges as a promising solution to remove barriers towards a carbon-free economy in Europe and beyond. Motivated by the recent developments in Denmark with the decision to construct the world's first artificial Offshore Energy Hub, this paper investigates how the lowest cost for green hydrogen can be achieved. A model proposing an integrated design of the hydrogen and offshore electric power infrastructure, determining the levelised costs of both hydrogen and electricity, is proposed. The economic feasibility of hydrogen production from Offshore Wind Power Hubs is evaluated considering the combination of different electrolyser placements, technologies and modes of operations. The results show that costs down to 2.4 €/kg can be achieved for green hydrogen production offshore, competitive with the hydrogen costs currently produced by natural gas. Moreover, a reduction of up to 13% of the cost of wind electricity is registered when an electrolyser is installed offshore shaving the peak loads.

**Abbreviations**

| | |
|---|---|
| AC | Alternate current |
| AEL | Alkaline electrolyser |
| DC | Direct current |
| HVDC | High voltage direct current |
| OWPP | Offshore wind power plant |
| PEMEL | Proton exchange membrane electrolyser |
| SOEL | Solid oxide electrolyser |
| VSC | Voltage source converter |

**Symbols**

| | |
|---|---|
| $A$ | Area, $m^2$ |
| $CapEx$ | Capital expenditures, M€ |
| $CF$ | Capacity factor, % |
| $CS$ | Cold start time, min |
| $D$ | Diameter, mm |
| $DR$ | Discount rate, % |
| $e$ | Specific energy, $kWh/m^3$ |
| $E$ | Energy, GWh |
| $f$ | Footprint, $m^2$ |
| $L$ | Length, km |
| $LCOE$ | Levelised cost of electricity, €/MWh |
| $LCOH$ | Levelised cost of the hydrogen, €/kg |
| $LT$ | Lifetime, - |
| $OpEx$ | Operational expenditures, M€/a |
| $OH$ | Operating hours, h |
| $N$ | Number, - |
| $\dot{m}$ | Mass flow rate, kg/h |
| $\tilde{m}$ | Molar mass, kmol/kg |
| $M$ | Annual mass, kg/a |
| $P$ | Power, GW |
| $p$ | Pressure, bar |
| $R$ | Ideal gas universal constant, kJ/kg/K |
| $RC$ | Reference cost, - |
| $RP$ | Reference power, MW |
| $RU$ | Reference unit, - |
| $SF$ | Scale factor, - |
| $t$ | Time, h |
| $W$ | Water consumption, l/kg |
| $T$ | Temperature, K |
| $\dot{V}$ | Volumetric flowrate, $m^3$/h |
| $V$ | Volume, $m^3$ |
| $\varphi$ | Power load, % |
| $\eta$ | Efficiency, % |

**Subscripts and superscripts**

| | |
|---|---|
| $COMP$ | Compressor |
| $EQ$ | Equipment |
| $ELEC$ | Electrolyser |
| $ELEN$ | Electrical energy |
| $DES$ | Desalination unit |
| $H$ | Hour |
| $HS$ | Hub-to-shore |
| $HUB$ | Hub |
| $H_2$ | Hydrogen |
| $IG$ | Inter-array grid |
| $IN$ | Inlet |
| $MAX$ | Maximum value |
| $MEAN$ | Mean value |
| *MIN* | Minimum value |
| *NEQ* | Non-equipment |
| *OUT* | Outlet |
| *PIPE* | Pipeline |
| *PS* | Protected shore |
| *RG* | Real gas |
| *S* | Section |
| *ST* | Station |
| *OWPP* | Offshore wind power plant |
| *WAT* | Water |
| *Y* | Year |

# 1 Introduction

## 1.1 Background

Concrete actions to accelerate the transition to a net-zero greenhouse gas emissions society have been taken across the European Union (EU) and beyond [1]. In February 2021, the Danish Parliament mandated the construction of the first artificial Energy Island in the North Sea as an initial step to harvest the abundant far offshore wind potential [2,3]. This Energy Island [4] will act as a Hub, interconnecting 3 GW of offshore wind power plants (OWPPs) and transmitting the produced electricity to shore, at much lower costs than OWPPs singularly connected to shore [5] (**Figure 1**).

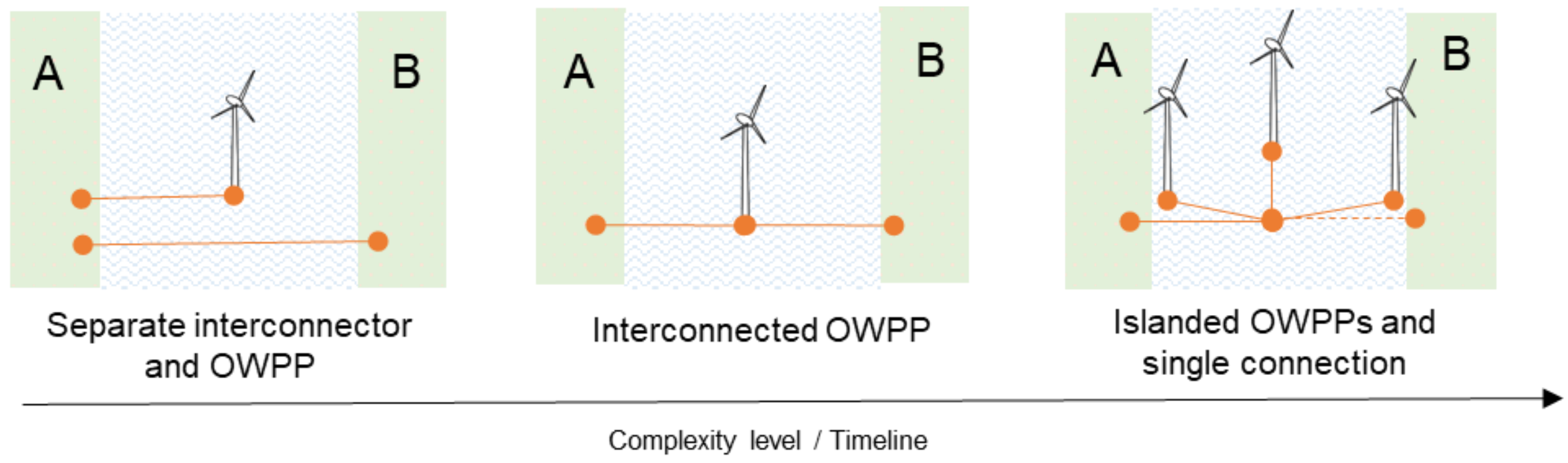

**Figure 1**. Evolution of the offshore power transmission infrastructure [6]. A, B: generic countries.

Only in the EU, the OWPPs capacity is expected to increase from the current 12 GW to 300 GW by 2050, of which 60 GW will be installed by 2030 [7,8]. International consortia, including countries surrounding the North Sea, are planning the next steps with the construction and future expansion of such offshore energy Hubs [9,10].

However, the integration of massive amounts of offshore wind introduces three main challenges. First, the high variability of wind power production places the supply-demand grid balance at risk. Second, the planned offshore installations require grid reinforcements in the order of billions of Euros [5,11]. Third, electricity will still face challenges with penetrating the so-called hard-to-abate sectors (e.g. heavy-duty road transport, aviation, shipping, and the steel industry), for which more energy-dense carriers are required.

Water electrolysis, using green electricity to generate hydrogen, is a potential solution to these challenges. Storable for longer periods and in larger quantities than electricity, hydrogen can support the supply-demand balance of the grid, help avoid grid reinforcements, and form the basis of green fuels (e.g. methane, ammonia, and methanol) [12]. Acknowledging these benefits, EU members set the ambitious goal to install electrolysers of 40 GW total capacity in Europe by 2030, and support the installation of an additional 40 GW in the EU's neighbourhood, to have this imported to the EU [13].

### 1.2 Cost of green hydrogen produced through offshore electrolysis

Despite the growing interest in hydrogen production, the literature regarding offshore electrolysis using electricity produced from offshore wind power is very limited. This is due to the cost of the electricity produced from offshore wind power parks, which has been higher than other renewable resources. Today, the declining costs and the large availability of offshore wind power makes this energy source a promising option for the large-scale production of hydrogen. On the other hand, offshore electrolysis has been seen as a promising solution to reduce the cost of the hydrogen delivered onshore and to minimise the investment in the electrical grid connecting the OWPPs to shore.

Meier [14] performed a cost analysis for hydrogen production on an offshore platform in Norway, through electrolysis powered by a 100 MW wind farm, resulting in a cost of production of 5.2 €/kg. Jepma and Van Schot [15] found that hydrogen, produced on existing oil and gas platforms, can have a cost of 2.84 €/kg, considering a future scenario accounting for the rapid expansion of the offshore wind energy capacity in the Dutch continental shelf of the North Sea, and also internalising the savings due to the avoided grid extensions. In the following study, Jepma et al. [16] calculated the cost of converting 100% of the power of a wind farm to hydrogen in the order of 2.50-3.50€/kg, using existing platforms and gas grids and total offshore conversion. If the saving obtained by the avoided extension of the electrical grid are internalised in the cost of the hydrogen, this would fall to 1-1.75€/kg.

Crivellari and Cozzani [17] presented an analysis of alternative power-to-gas and power-to-liquid strategies for the conversion of offshore wind power into different chemical energy vectors. The study showed that gaseous hydrogen produced offshore and transmitted through a new pipeline is the most expensive among the other alternatives, with a cost of 212 €/MWh (equivalent to 6.4 €/kg), but it presents the best performance in terms of $CO_2$ equivalent emissions.

To date, the cost of producing both hydrogen and electricity from a multi-GW offshore energy Hub, comprising multiple OWPPs, has not been assessed, and alternative topologies regarding the integration of electrical and hydrogen infrastructure have not been explored.

## 1.3 Motivation and objectives

Considering that the production of green hydrogen will be closely associated with the Offshore Energy Hubs, and the central role hydrogen is expected to play in the energy economy, one key question arises: how can we achieve the lowest cost for green hydrogen delivered onshore?

To answer this question, this paper presents a holistic approach, proposing a techno-economic model which considers the complementary design of both hydrogen and offshore electric power infrastructure, so far considered only separately [14–17]. Our approach allows us to identify the interactions and potential synergies between the two energy carriers, and determine the levelised cost of hydrogen (LCOH) and electricity (LCOE). Our analyses consider, among others, three main parameters:

- the placement of the electrolyser: onshore, offshore or in-turbine;
- the share of the electricity routed towards hydrogen production: "hydrogen-driven", if priority is given to the electrolysers, or "electricity-driven", if only the excess electricity is directed to the electrolysers;
- the type of electrolyser technology: alkaline, proton exchange membrane, or solid oxide.

The reference values for the calculated LCOH are the cost of grey and blue hydrogen. Grey hydrogen, produced from natural gas, costs 0.8-2.7 €/kg [18]; blue hydrogen, produced from natural gas as well but also including the carbon capture, costs 1.3-2.4 €/kg [18]. The calculated LCOE is compared with the current cost of offshore wind electricity in Europe, which is 45-79 €/MWh [19].

## 1.4 Case study and applicability to other regions

The Hub and Spoke (H&S) configuration is a recently explored grid connection system. This envisions the deployment of an offshore Hub, where AC-electricity from surrounding offshore wind power parks (OWPPs) is converted to DC, and then transported onshore via HVDC. For far OWPPs, the H&S concept has been found more cost-effective than the radial HVDC connections to individual wind OWPPs, benefiting from the economies of scale of collecting a large amount of power [20]. This study proposes a reference case of a 12 GW Hub, as assumed by the North Sea Power Hub Consortium's work [20], located 380 km from Esbjerg (Denmark) [21] (**Figure 2**). Although applied to a 12 GW Energy Island in the North Sea, the same objective of this study can be reached for other regional contexts and different sizes, with the same methodological process.

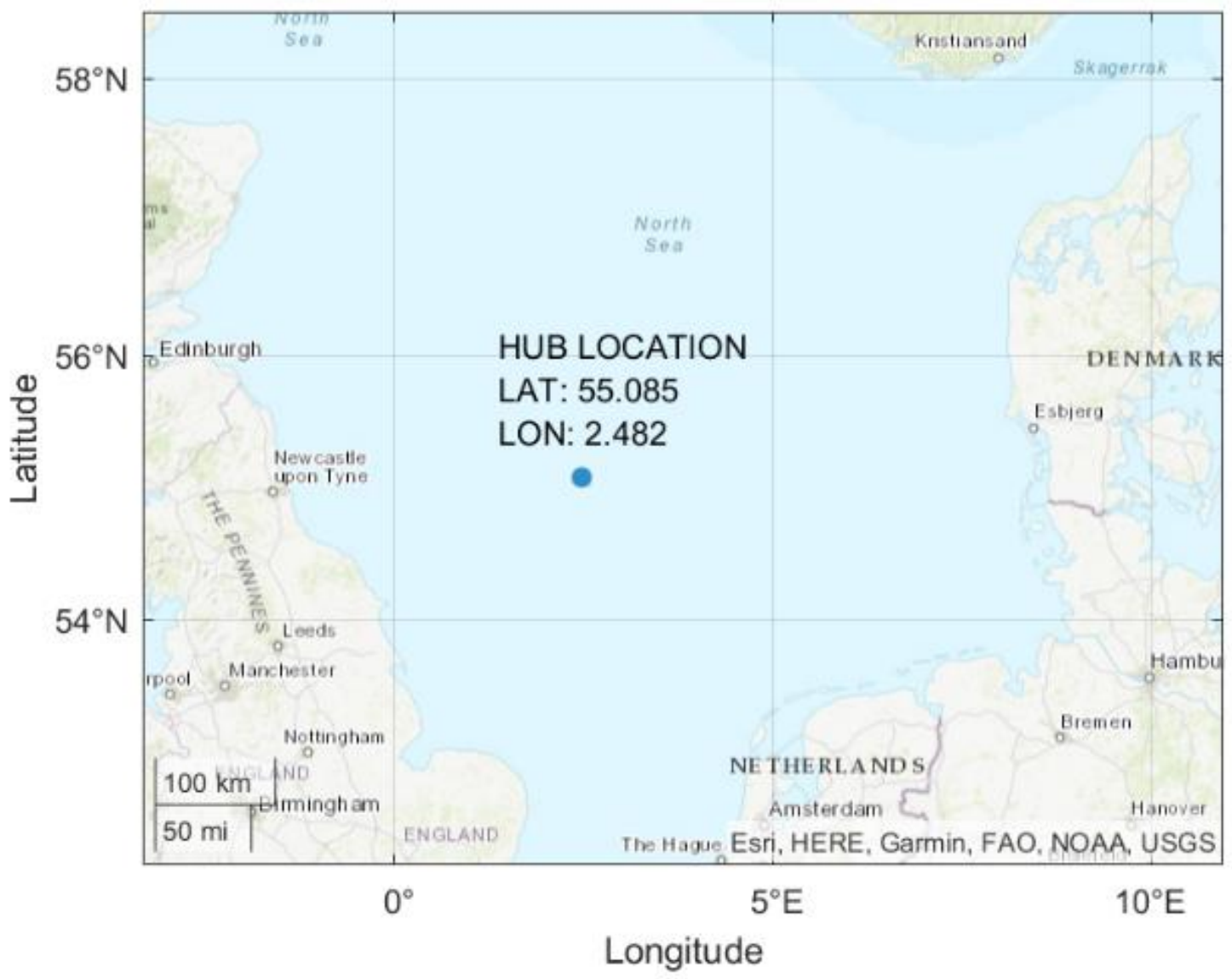

**Figure 2**. Original Hub position in the North Sea [21].

## 2 Methodology

In **Section 2.1**, the alternative placements for the electrolyser are described, characterising the offshore infrastructure necessary In **Section 2.2**, the relation between the share of the electricity converted into hydrogen and the share of the electricity delivered onshore is defined, characterising the operations of the electrolyser. In **Section 2.3**, the units of the equipment involved in the electrical and hydrogen infrastructures are modelled, calculating the mass and energy balances between them, defining their sizes. In **Section 2.4**, the techno-economic assessment of the alternative placements configuration is carried out, considering the calculated sizes of the equipment. The final result provides the LCOH and the LCOE delivered onshore used to compare the different scenarios. The model used is built in Matlab 2019b [22] and Cantera 2.4 [23].

### 2.1 Electrolyser placement

Three different electrolyser placements (**Figure 3**), along with their related infrastructures, have been investigated:

I. Onshore: the electricity produced by all OWPPs is collected at the Hub and transmitted to shore, where hydrogen is produced by a single electrolyser, then compressed to grid pressure.

II. Offshore: the electricity produced by all OWPPs is transmitted to the Hub, where hydrogen is produced by a single electrolyser, using desalinated seawater, then compressed and transported to shore via pipeline.

III. In-turbine: the electrolysers, paired with desalination units, are located inside or next to the tower of each wind turbine (WT). The produced hydrogen is transported to the Hub via pipelines that connect groups of WTs. On the Hub, the hydrogen is collected, compressed, and transported to shore via a pipeline.

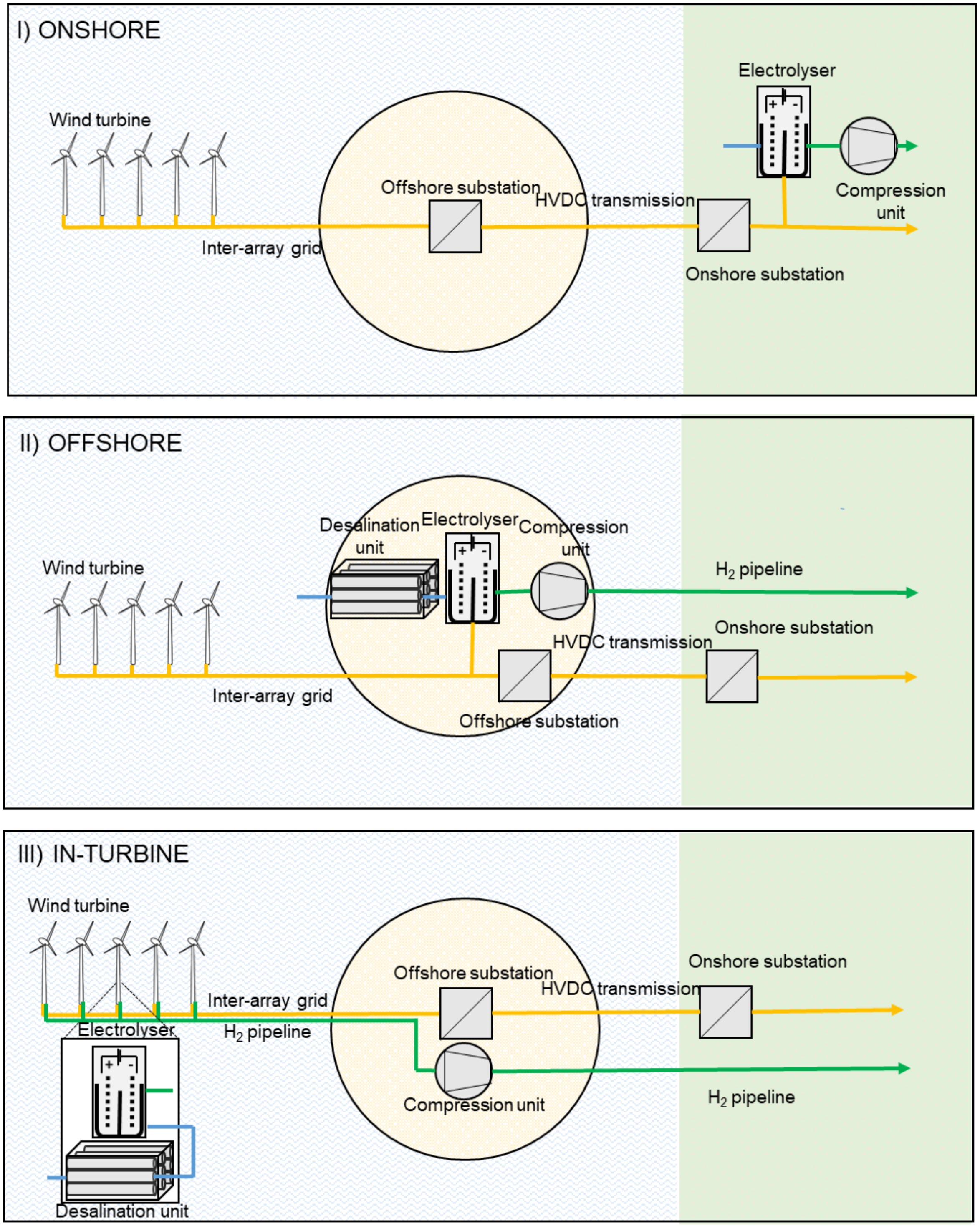

**Figure 3**. Schematic representation of the electrolyser placements.

**Figure 4** shows the flowchart of the configurations: onshore, offshore and in-turbine. The placement of the electrolyser determines the section of the offshore power system at which the electricity is used, identified by the subscripts *I*, if in-turbine, *II*, if on the offshore Hub, *III* if onshore.

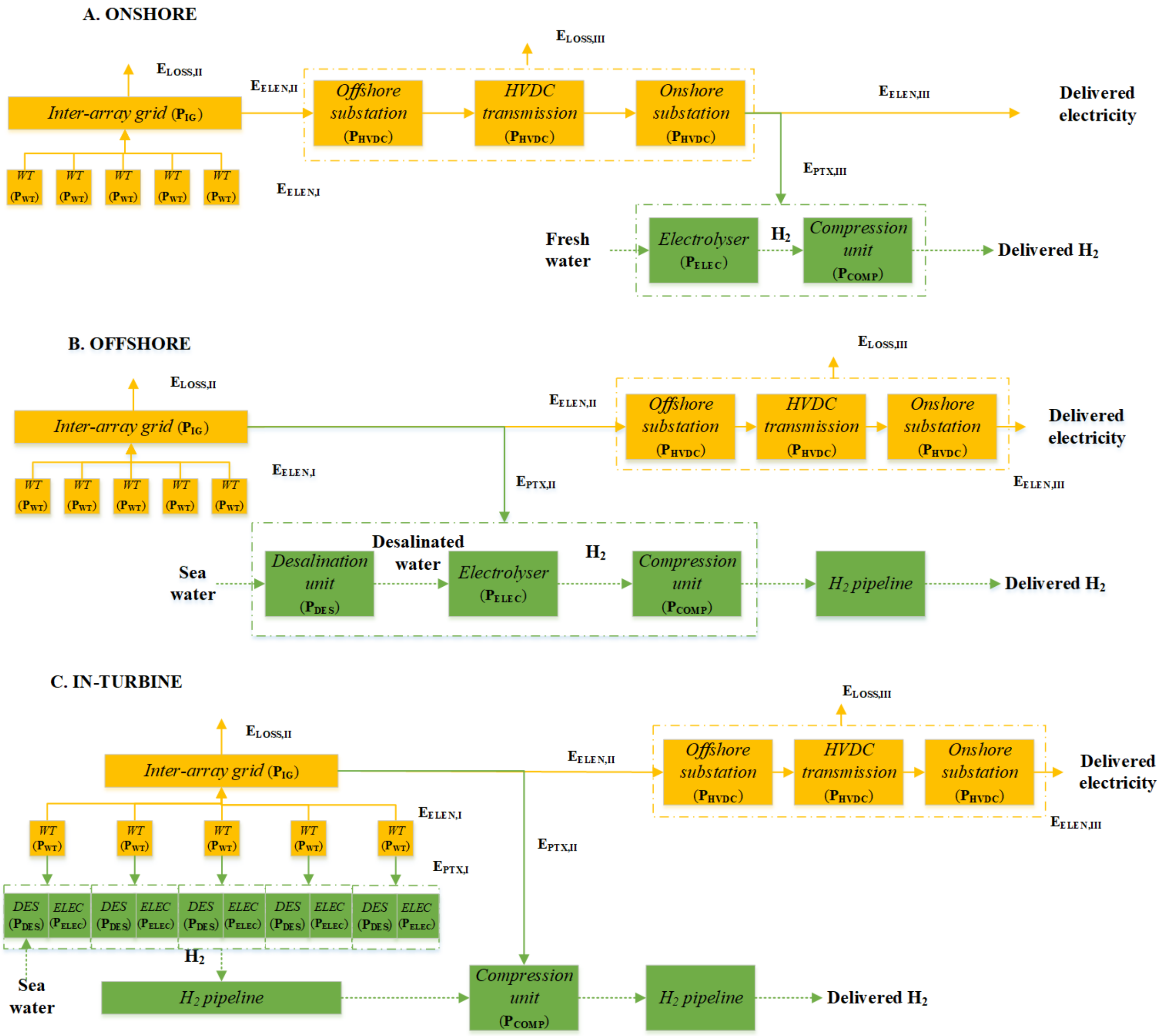

**Figure 4**. Flowchart of the three alternatives. Notes: only one group of WTs connected to the inter-array grid is represented, $E_{ELEN}$ represents the total energy at each section. WT: wind turbine; DES: desalination unit; ELEC: electrolyser.

## 2.2 Electricity and hydrogen co-generation

When co-generating electricity and hydrogen, two opposite operation modes can be envisioned and they are described as follows:

I. hydrogen-driven: the electricity generated by the Hub firstly covers the nominal electrolyser capacity, while the remaining electricity is directed to shore. In this case, the electrolyser uses the base load electricity production.

II. Electricity-driven: the electrolyser uses only the excess electricity generated. In this case, priority is given to covering the electricity demand, and the electrolyser shaves the peak load.

These two alternative operation modes define different electrical energy input for an electrolyser, due to the availability of the energy generated by the Hub. Hydrogen-driven operations ensure higher utilisation of the electrolyser, due to a more frequent electrical energy input, compared to electricity-driven operations, which rely on less frequent peaks of energy production. An example of the effect of these two types of operation on the electrolyser utilisation is represented in **Figure 5**.

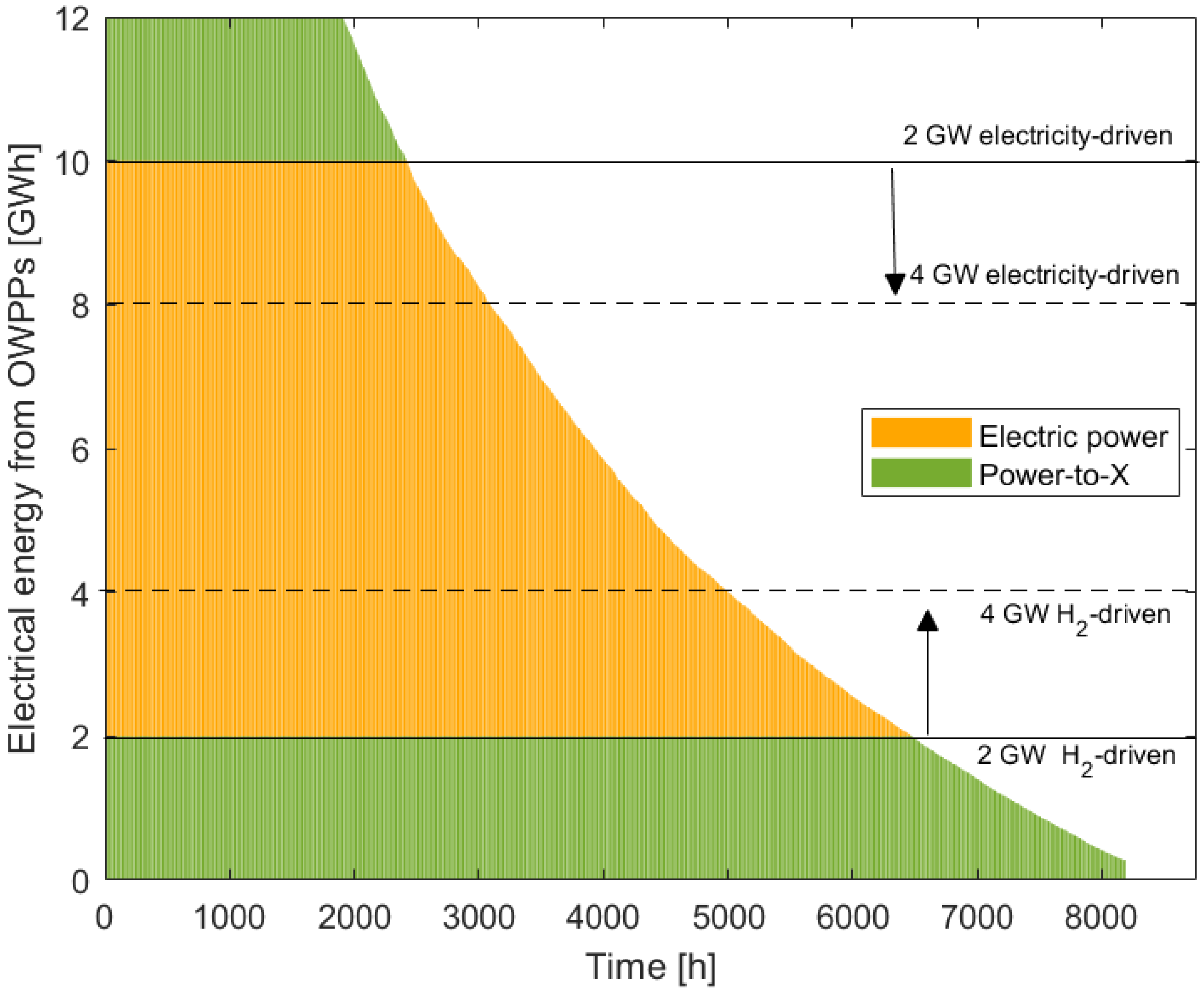

**Figure 5.** Duration curve of a 12 GW Hub, illustrating the hydrogen-driven and the electricity-driven operations. The green area identifies the electrical energy input of the electrolyser, in this example of 2 GW, for the two operation types. Dashed lines delimit the electrical energy input for a 4 GW electrolyser; the orange area enclosed by the solid and dashed line represents the difference in the electrical energy input between a 2 GW and a 4 GW electrolyser in the same operation mode.

The rate of utilisation of the electrolyser impacts on the cost of the hydrogen produced. Therefore, this study analyses these two types of operations and the whole range of possible hydrogen to electricity ratios: from 100% electricity and 0% hydrogen (no electrolyser installed) to 0% electricity and 100% hydrogen (or 12 GW electrolyser installed over a 12 GW Hub), resizing the electricity and hydrogen infrastructures accordingly in each case.

The electric energy used for hydrogen at the section $S$, $E_{PTX,S}(t)$, is calculated using Eq. (1).

$$E_{PTX,S}(t) = \begin{cases} min\left(P_{ELEC} \cdot \Delta t, E_{HUB}(t) - \sum_{i=I}^{S} E_{LOSS,i}(t)\right), & if\ hydrogen-driven \\ E_{HUB}(t) - \sum_{i=I}^{S} E_{LOSS,i}(t) - min\left(P_{HUB} \cdot \Delta t - P_{ELEC} \cdot \Delta t, E_{HUB}(t) - \sum_{i=I}^{S} E_{LOSS,i}(t)\right), & if\ electricity-driven \end{cases} \quad (1)$$

where $E_{HUB}(t)$ is the electricity generated by the entire Hub; $P_{ELEC}$ is the nominal capacity of the installed electrolyser; $E_{LOSS}(t)$ is the sum of all the electric losses upstream of the electrolyser located at location $S$.

Alternatively, the remaining electricity at the section $S$, $E_{ELEN,S}(t)$, is calculated using Eq. (2).

$$E_{ELEN,S}(t) = \begin{cases} E_{HUB}(t) - \sum_{i=I}^{S} E_{LOSS,i}(t) - min\left(P_{ELEC} \cdot \Delta t, E_{HUB}(t) - \sum_{i=I}^{S} E_{LOSS,i}(t)\right), & if\ hydrogen-driven \\ min\left(P_{HUB} \cdot \Delta t - P_{ELEC} \cdot \Delta t, E_{HUB}(t) - \sum_{i=I}^{S} E_{LOSS,i}(t)\right), & if\ electricity-driven \end{cases} \quad (2)$$

### 2.3 Process design model

Each technological unit is characterised in the following subsections, in which the main parameters of each technology are characterised, including their energy usage, $E$, and power, $P$, in units of gigawatt-hours and gigawatts, respectively.

#### 2.3.1 Offshore wind turbines

The Hub is assumed to be composed of wind turbines (WTs) of capacity 15 MW [20], $P_{WT}$. A representative hourly wind power yield is generated from the hourly wind speed data from the Copernicus ERA5 Dataset [24], for the analysed location, and considering the International Energy Agency's specifications for a reference 15 MW turbine [25]. The summed hourly electricity production of each WT for an entire year is identified as $E_{HUB}(t)$.

### 2.3.2 Inter-array grid

The Hub is assumed to be constituted by a series of concentrically distributed OWPPs of 1 GW each. Each group of 5 WT, $N_{WT}$, is then connected to the Hub through 66 kV AC cables [20]. The length of each string, $L_{IG}$, is calculated as the sum of the distance between each WT, $L_{WT}$, and the average distance of each OWPP to the Hub, $L_{HUB}$, using Eq. (3), in units of kilometres.

$$L_{IG} = L_{WT} \cdot (N_{WT} - 1) + L_{HUB} \tag{3}$$

The distance between each WT, $L_{WT}$, is calculated using Eq. (4), in units of kilometres.

$$L_{WT} = \sqrt[2]{\frac{P_{WT}}{PY_{WT}}} \tag{4}$$

where $PY_{WT}$ is the power yield of the WT, assumed 4.5 MW/km$^2$ [26].

The average distance of each $n$ OWPP from the Hub, $L_{HUB}$, is calculated using Eq. (5) in units of kilometres.

$$L_{HUB,OWPP(n)} = \frac{1}{2} \cdot \left| \left\lfloor \frac{n}{4} \right\rfloor \cdot \sqrt[2]{\frac{P_{OWPP}}{PY_{WT}} \cdot \frac{12}{\pi}} - \sqrt[2]{\frac{A_{OWPP} \cdot 4}{\pi}} \right| + \left\lfloor \frac{n}{4} \right\rfloor \cdot \sqrt[2]{\frac{P_{OWPP}}{PY_{WT}} \cdot \frac{4}{\pi}} \tag{5}$$

The nominal power of each string, $\bar{P}_{IG}$, is calculated using Eq. (6).

$$\bar{P}_{IG} = \frac{\bar{P}_{HUB} - \bar{P}_{ELEC,I}}{N_{IG}} \tag{6}$$

where $\bar{P}_{ELEC,I}$ is the electrolyser total installed capacity at the location $I$, of the infrastructure (i.e. in-turbine), equal to zero if the electrolyser is located elsewhere. $N_{IG}$ is the number of strings of the inter-array grid, calculated using Eq. (7).

$$N_{IG} = \frac{P_{HUB}}{P_{WT} \cdot N_{WT}} \tag{7}$$

The electricity loss in the inter-array grid, $E_{LOSS,II}(t)$, is calculated using Eq. (8).

$$E_{LOSS,II}(t) = N_{WT} \cdot E_{ELEN,I}(t) \cdot \eta_{IG} \tag{8}$$

where $\eta_{IG}$ is the coefficient of electric energy loss in the inter-array grid, assumed equal to 0.55% of the electric energy transmitted [27].

### 2.3.3 HVDC transmission

The electric energy generated is collected on the Hub, on which the AC is converted into high voltage direct current (HVDC) through voltage source converters.

The rated power for the offshore substation, HVDC cable and onshore substation, $P_{HVDC}$, is the difference between the total power of the Hub, $P_{HVDC}$, and the total power capacity of the electrolyser, $P_{ELEC}$, if installed in-turbine or offshore.

The electric energy loss in the HVDC transmission, $E_{LOSS,III}$, is calculated using Eq. (9).

$$E_{LOSS,III}(t) = E_{ELEN,II}(t) \cdot \left(\eta_{ST} \cdot N_{HVDC,ST} + \eta_{HS} \cdot L_{HS}\right) \tag{9}$$

where $\eta_{ST}$ is the energy loss at the conversion station; $N_{HVDC,ST}$ is the number of the substation, equal to 2; $\eta_{HS}$ is the energy loss per km; $L_{HS}$ is the distance from the hub to the shore. In this case, $\eta_{ST}$ is assumed to be equal to 1% [28]; $\eta_{HS}$ is assumed to be 0.0035% [28]; $L_{HS}$ is estimated to be 380 km from the Hub location to the Denmark shore[21].

### 2.3.4 Electrolyser system

In this work, the three main types of electrolysers are analysed, whose operational parameters used in the model for the main electrolyser technologies are listed in **Table 1**.

**Table 1**. Electrolyser system operational parameters for Alkaline (AEL), Proton Exchange Membrane (PEMEL) and Solid Oxide Electrolyser (SOEL).

| | AEL | PEMEL | SOEL | Ref. |
|---|---|---|---|---|
| Operating pressure, $p_{ELEC}$ [bar] | 30 | 55 | 5 | [29] [30] |
| Operating temperature, $T_{ELEC}$ [°C] | 80 | 85 | 675 | [30] |
| System electrical efficiency*, $\bar{\eta}_{ELEC}$ [%] | 66 | 62 | 79 | [30] |
| Stack operating time, $OH_{MAX}$ [h] | 82,500 | 85,000 | 61,320 | [31] [30] |
| Load range, $\varphi_{MIN}$ -$\varphi_{MAX}$ [% $\bar{P}_{ELEC}$] | 10-110 | 0-160 | 0-100 | [29] [30] |
| Cold start up (after 24h stop), $CS$ [min] | 20 | 5 | 60 | [32] [30] |
| Degradation, $\eta_{DEG}$ [%/1,000 h] | 0.10 | 0.10 | 0.50 | [33] |
| Plant footprint, $f_{ELEC}$ [$m^2$/GW] | 95,000 | 48,000 | 7,000 | [29] [30] |

*On a lower heating value (LHV) basis; including the energy consumption of the electrolyser stacks, gas water separators, demisters, gas drying, water management, lye system (for AEL), system control, power supply [32].

The chemical energy of the hydrogen produced, $E_{H_2}(t)$, is calculated using Eq. (10).

$$E_{H_2}(t) = \begin{cases} E_{ELEC}(t) \cdot \eta_{ELEC}(t) \cdot \left(1 - \frac{CS}{60} \cdot \left\lfloor \frac{\sum_1^{24} 1 - B(t-i)}{24} \right\rfloor \right), & P_{ELEC} \cdot \Delta t \cdot \varphi_{MIN} \leq E_{ELEC}(t) < P_{ELEC} \cdot \Delta t \\ 0, & E_{ELEC}(t) < \bar{P}_{ELEC} \cdot \Delta t \cdot \varphi_{MIN} \end{cases} \tag{10}$$

where $\bar{P}_{ELEC}$ is the nominal capacity of the electrolyser; $\varphi_{MIN}$ is the minimum partial capacity of the electrolyser; $CS$ is the cold start time in units of minutes; $B$ is a Boolean parameter, whose value indicates the operation of the electrolyser at the hour $t$-$i$, where $B(t-i) = 1$ if $E_{HYD}(t-i) > 0$ (or the electrolyser is on), or $B(t-i) = 0$ (or the electrolyser is off) otherwise. If for consecutive 24 hours the electrolyser is not operational, a cold start is assumed to be necessary for the operational hour $t$.

$\eta_{ELEC}(t)$ is the efficiency of the electrolyser at the hour $t$. For $\eta_{ELEC}(1)$ the nominal efficiency is considered (**Table 1**). The effect of the efficiency degradation is calculated using Eq. (11).

$$\eta_{ELEC}(t+1) = \eta_{ELEC}(t) \cdot \left(1 - \frac{\eta_{DEG}}{1{,}000} \cdot B(t)\right) \tag{11}$$

$\eta_{DEG}$ is the degradation of the efficiency (**Table 1**). The number of operational hours of the electrolyser is calculated using Eq. (12).

$$OH = \sum_{t=1}^{LT_H} B(t) \tag{12}$$

where $LT_H$ is the lifetime of the plant in hours, in this case, assumed equal to 8,760 per year for 30 years. During the lifetime of the plant, the stack is replaced when $t = n \cdot OH_{MAX}$ for each $n$. Consequently, for $\eta_{EL}(n \cdot \mathrm{OH}_{MAX} + 1)$, the nominal efficiency, $\bar{\eta}_{ELEC}$, is restored.

The capacity factor of the electrolyser, $CF_{EL}$, defined as the share of operating hours of the electrolyser during its lifetime, is calculated using Eq. (13).

$$CF_{H_2} = \frac{\sum_{t=1}^{LT_H} E_{ELEC}(t) \cdot B(t)}{P_{ELEC} \cdot \Delta t \cdot LT_H} \tag{13}$$

The hydrogen mass flow rate is calculated using Eq. (14), in units of kilograms per hour.

$$\dot{m}_{H_2}(t) = \frac{E_{H_2}(t) \cdot 10^6}{LHV_{H_2}} \tag{14}$$

where $LHV_{H_2}$ is the lower heating value of the hydrogen, equal to 33.3 kWh/kg.

#### 2.3.5 Desalination unit

If offshore or in-turbine, the water for the electrolyser system shall be supplied by a desalination unit. In this analysis, it is assumed that the desalination unit is based on reverse osmosis. The volumetric flow rate of the water is calculated using Eq. (15), in units of cubic metres per hour.

$$\dot{V}_{H_2O}(t) = \dot{m}_{H_2}(t) \cdot W_{DES} \cdot 10^{-3} \tag{15}$$

where $W_{DES}$ is the water consumption for each kilogram of hydrogen produced, assumed to be 15 litres of water per kilogram of hydrogen [34]. The nominal volumetric flow rate of the desalination unit, $\bar{V}_{DES}$, is assumed to be the maximum value of $\dot{V}_{H_2O}(t)$.

The electric energy consumption of the desalination unit is calculated using Eq. (16).

$$E_{DES}(t) = \dot{V}_{H_2O}(t) \cdot e_{DES} \cdot 10^{-6} \tag{16}$$

where $e_{DES}$ is the energy consumption per cubic meter of water processed, assumed to be 3.5 kWh $m^{-3}$ [29].

2.3.6 Compression unit

The hydrogen produced is compressed into a pipeline. The formula for adiabatic compression [35], Eq. (17), is used to calculate the required energy, $E_{COMP}(t)$.

$$E_{COMP}(t) = \frac{286.76 \cdot \dot{m}_{H_2}(t) \cdot T_{MEAN}}{\eta_{COMP} \cdot G_{H_2} \cdot 3.6 \cdot 10^9} \cdot \left(\frac{\gamma \cdot N_{ST}}{\gamma - 1}\right) \cdot \left[\left(\frac{p_{COMP,OUT}}{p_{COMP,IN}}\right)^{\frac{\gamma-1}{\gamma \cdot N_{ST}}} - 1\right] \cdot \Delta t \tag{17}$$

where $\eta_{COMP}$ is the compression efficiency, assumed to be 50% [34] due to frequent load variations; $\gamma$ is the ratio between the specific heat capacities for hydrogen ($\gamma = cp/cv$); $N_{ST}$ is the number of compression stages, for simplicity assumed as 1; $G_{H_2}$ is the gas gravity of the hydrogen, 0.0696, defined as the molar mass of hydrogen divided by the molar mass of air; $T_{MEAN}$ is the mean temperature, assumed to be 285.15 K [36].

The three placements of the electrolyser determine the value of $p_{COMP,IN}$ and $p_{COMP,OUT}$:

I. Onshore: $p_{COMP,IN} = p_{ELEC}$ (**Table 1**), $p_{COMP,OUT} = p_{TRANS}$ (assumed to be 70 bar [16]).

II. Offshore: $p_{COMP,IN} = p_{ELEC}$ (**Table 1**), $p_{COMP,OUT} = p_{PIPE,IN}$.

III. In-turbine: $p_{COMP,IN} = p_{PIPE,OUT}$, as outlet pressure of the pipeline connecting the string of WTs to the Hub, $p_{COMP,OUT} = p_{PIPE,IN}$, as the inlet pressure of the pipeline connecting the Hub to shore.

The values of $p_{PIPE,OUT}$ and $p_{PIPE,IN}$ are determined in the following subsection.

The nominal power of the compressor, $\bar{P}_{COMP}$, is assumed to be the maximum value of $E_{COMP}(t)$ per hour.

2.3.7 Hydrogen pipeline

The sizes of the pipelines, from the WTs to the Hub and from the Hub to shore, are determined using Eq. (18) [35]

$$\dot{V}_{H_2}(T_b, p_b) = \frac{1.1494}{24} \cdot (10^{-3}) \cdot \left(\frac{T_b}{p_b}\right) \cdot \sqrt[2]{\frac{D^5 \cdot \left(p_{PIPE,IN}^2 - p_{PIPE,OUT}^2\right)}{Z_{MEAN} \cdot T_{MEAN} \cdot G_{H_2} \cdot L \cdot \lambda}} \tag{18}$$

where $\dot{V}_{H_2}(T_b, p_b)$ is the volumetric flowrate of the hydrogen at standard conditions ($T_b$ =288.15 K, $P_b$ = 1 bar [35]), in units of cubic metres per hour; $p_{PIPE,IN}$ and $p_{PIPE,OUT}$ are the upstream and downstream pipeline pressures in units of kilopascals; $Z_{MEAN}$ is the dimensionless compressibility factor; $\lambda$ is the dimensionless coefficient of friction; $L$ is the length of the pipeline in units of kilometres; $D$ is the inner diameter of the pipeline in units of metres.

Pipelines from the OWPPs to the Hub and from the Hub to shore are deployed, having the following values:

I. For the pipelines from the OWPPS to the Hub: $L = L_{IG}$, $p_{PIPE,IN} = p_{ELEC}$

II. For the pipelines from the Hub to shore: $L = L_{HS}$, $p_{PIPE,OUT}$ = 70 bar

See Appendix B for further details on the pipeline sizing.

### 2.3.8 Artificial island

The Hub hosting the offshore equipment is assumed to be a sand island, as this is considered to be more cost-effective than other types of offshore platforms in the case of a large hub in shallow waters [9]. The Hub shall have a surface, $A_{HUB}$, able to host the HVDC offshore substation and the electrolyser. $A_{HUB}$ is calculated using Eq. (19), in units of square metres.

$$A_{HUB} = P_{HVDC} \cdot f_{HVDC} + P_{ELEC} \cdot f_{ELEC} \tag{19}$$

where $f_{HVDC}$ is the footprint of the offshore substation, here assumed to be 4’860 $m^2$ /GW [37], and $f_{ELEC}$ is the footprint of the electrolyser. The volume of the sand used to build the island, $V_{HUB}$, and the area of the shoreline assumed to be protected, $A_{PS}$, are simplified considering the island has the shape of a truncated cone.

The volume of the hub, $V_{HUB}$, is calculated using Eq. (20).

$$V_{HUB} = \frac{1}{3} \cdot s \cdot \pi \cdot ({r_{SB}}^3 - {r_{HUB}}^3) \tag{20}$$

where $r_{HUB}$ is the radius at the surface level, and $r_{SB}$ is the radius at the seabed level, in units of metres, calculated using Eq. (21) and Eq. (22), respectively.

$$r_{HUB} = \sqrt{\frac{A_{HUB}}{\pi}} \tag{21}$$

$$r_{SB} = r_{HUB} + h/s \tag{22}$$

where $s$ is the slope of the truncated cone, assumed to be 75%; $h$ is the depth of the seabed, assumed 30 m [20] to which is added 10% of elevation, to be over the sea level.

Moreover, the area of the shoreline assumed to be protected, $A_{PS}$, in units of square metres, is calculated, using Eq. (23).

$$A_{PS} = \pi \cdot {r_{SB}}^2 + \pi \cdot r_{SB} \cdot \sqrt{{r_{SB}}^2 \cdot (1+s^2)} - \pi \cdot {r_{HUB}}^2 - \pi \cdot r_{HUB} \cdot \sqrt{{r_{HUB}}^2 \cdot (1+s^2)} \quad (23)$$

## 2.4 Techno-economic analysis

The LCOE and the LCOH are used to compare the alternative configurations and calculated as shown in **Table 2**. The LCOE at each section of the electric power infrastructure is calculated using Eq. (24)-(26) and expressed in units of Euro per megawatt-hour of electricity. The LCOH is calculated using Eq. (27) and expressed in units of Euro per kilogram of hydrogen produced.

**Table 2**. Levelised cost of the energy and levelised cost of the hydrogen equations. Note: $LCOE_{III}$ is also the final cost of the electricity delivered onshore.

| Symbol | Value | Eq. |
|---|---|---|
| $LCOE_I$ | $\sum_{Y=0}^{LT_Y} \frac{CapEx_{ELEN,I,Y} + OpEx_{ELEN,I,Y}}{(1+DR)^Y} \Big/ \sum_{y=0}^{LT_Y} \frac{E_{HUB,Y}}{(1+DR)^Y}$ | (24) |
| $LCOE_{II}$ | $\sum_{y=0}^{LT_Y} \frac{LCOE_I \cdot E_{ELEN,I,Y} + CapEx_{ELEN,II,Y} + OpEx_{ELEN,II,Y}}{(1+DR)^Y} \Big/ \sum_{Y=0}^{LT_Y} \frac{E_{ELEN,I,Y} - E_{LOSS,II,Y}}{(1+DR)^Y}$ | (25) |
| $LCOE_{III}$ | $\sum_{Y=0}^{LT_Y} \frac{LCOE_{II} \cdot E_{ELEN,II,Y} + CapEx_{ELEN,III,Y} + OpEx_{ELEN,III,Y}}{(1+DR)^Y} \Big/ \sum_{Y=0}^{LT} \frac{E_{ELEN,II,Y} - E_{LOSS,III,Y}}{(1+DR)^Y}$ | (26) |
| $LCOH$ | $\sum_{Y=0}^{LT_Y} \frac{LCOE_s \cdot E_{PTX,S,Y} + CapEx_{PTX,Y} + OpEx_{PTX,Y}}{(1+DR)^Y} \Big/ \sum_{Y=0}^{LT_Y} \frac{M_{H_2,Y}}{(1+DR)^Y}$ | (27) |

$DR$ is the discount rate, which reflects the financial return and the project risk, here assumed to be 5% [16]; $LT_Y$ is the lifetime of the project as the lifetime of the system, 30 years [20]; $E_{ELEN}$, $E_{LOSS}$ and $M_{H_2}$ are the electric energy, energy loss and mass of hydrogen cumulated over the year $Y$.

CapEx and OpEx are the sum of the CapEx and OpEx of each component deployed in the electric and PtX infrastructure in the year $Y$. See Appendix B for the details of the costs of each component.

It is important to notice that the electric energy used for hydrogen production, $E_{PTX,S}$, is considered to have a cost equal to the $LCOE_S$ calculated at the location $S$ of the electric power infrastructure where the electrolyser, desalination unit and compression unit are located.

## 3 Results and discussion

### 3.1 Electrolyser technology comparison.

Among the three electrolyser technologies, AEL presents the lowest LCOH, due to a better trade-off between costs and operational parameters, but with only negligible differences in comparison to PEMEL and SOEL. A significant difference in the LCOHs is observed in the in-turbine placement, in which SOEL register a higher LCOH. The full LCOH comparison between the three technologies and visualisations are provided in the Supplementary Material.

This is due to the combined effects of higher CapEx for small sizes, due to economies of scale, and lower operating pressure, which requires the use of external additional compression, absorbing part of the electric energy directed to hydrogen production, thus decreasing its hydrogen production. Another major weakness of SOEL is the higher degradation rate of its stack, which leads to a more frequent replacement compared to the other two technologies. Therefore, despite the higher efficiency, the LCOH for SOEL is greater compared to the other two technologies.

Inversely, PEMEL achieves higher capacity factors (CFs) mainly due to its lower electric consumption: PEMEL operates at a higher pressure, 55 bar [29], which limits the use of an external compression unit. This allows a higher share of electricity to be used for hydrogen production compared to AEL and SOEL.

AEL's lower LCOH compared with PEMEL and SOEL transcends across all operating and placement scenarios in this article. Therefore, for the sake of readability, the next sections refer only to the results associated with AEL, while the results for each type of electrolyser can be found in the Supplementary Material.

### 3.2 Hydrogen-driven operation mode

The main results for the hydrogen-driven operation are presented in **Figure 6**. Three main factors affect the LCOH: the utilisation of the infrastructure, the cost of the electricity supplied to the electrolyser, and the economies of scale for the different components.

The utilisation of hydrogen or electricity infrastructures can be described by their CF. For the hydrogen-driven operation, the larger the electrolyser installed capacity, the lower its CF (**Figure 5**). The CF is affected in two ways by the placement of the electrolyser. On one side, the lower the electrical consumption of the ancillary equipment associated with that placement is (i.e. desalination and compression units), the more electricity is used by the electrolyser, increasing its CF. On the other side, the more components of the electricity infrastructure upstream of the electrolyser are, the larger the electrical losses are, reducing the CF of the electrolyser.

A second driver for the LCOH is the cost of the electricity supplied to the electrolyser. The use (or not) of the inter-array grid and HVDC infrastructure, along with the associated electricity losses, to transport electricity to the electrolyser determines the cost of the electricity used. Therefore, the cost of electricity used by the electrolysers placed onshore is higher than the cost of electricity used by same-sized electrolysers placed offshore, not using the offshore HVDC infrastructure, or in-turbine, not even using an inter-array grid infrastructure.

The third driver is associated with the economies of scale related to the electrolysers and the pipelines. Because of their modularity, the economies of scale of the electrolyser are evident only for sizes lower than 100 MW, becoming highly noticeable for sizes lower than 10 MW[33]. Therefore, in-turbine electrolysers, having capacities necessarily below the size of the WT (15 MW), are affected more strongly by economies of scale.

In the case of pipelines of the same length, increasing the diameter, the cost per capacity decreases. Therefore, in the case of small scales in-turbine placement, the LCOH is penalised by a large number of pipelines from the OWPPs to the Hub.

As shown in **Figure 6.A**, placing the electrolyser on the Hub achieves the lowest LCOH, with a minimum of 2.4 €/kg. Irrespective of the installed capacity, offshore electrolysis can produce hydrogen at a cost-competitive with the grey hydrogen. **Figure 6.B** shows how LCOE varies with different electrolyser placements and installed capacity. In the case of hydrogen-driven operation, as the electrolyser capacity increases, the utilisation of the electricity infrastructure (i.e. $\mathrm{CF}\text{-}El$) decreases; this results to LCOE increasing when large amounts of hydrogen are produced, in case the electrolyser is placed offshore or in-turbine. Besides the lower CF, the main driver for this increase is the fixed costs of HVDC and inter-array cables, which heavily depend on their length and significantly less on their capacity. Moreover, as expected, in case the electrolyser is placed onshore, the LCOE remains unaffected. For the largest part of installed electrolyser capacities, LCOE remains widely competitive with current offshore wind installations, with the lowest LCOE estimated at 45 €/MWh.

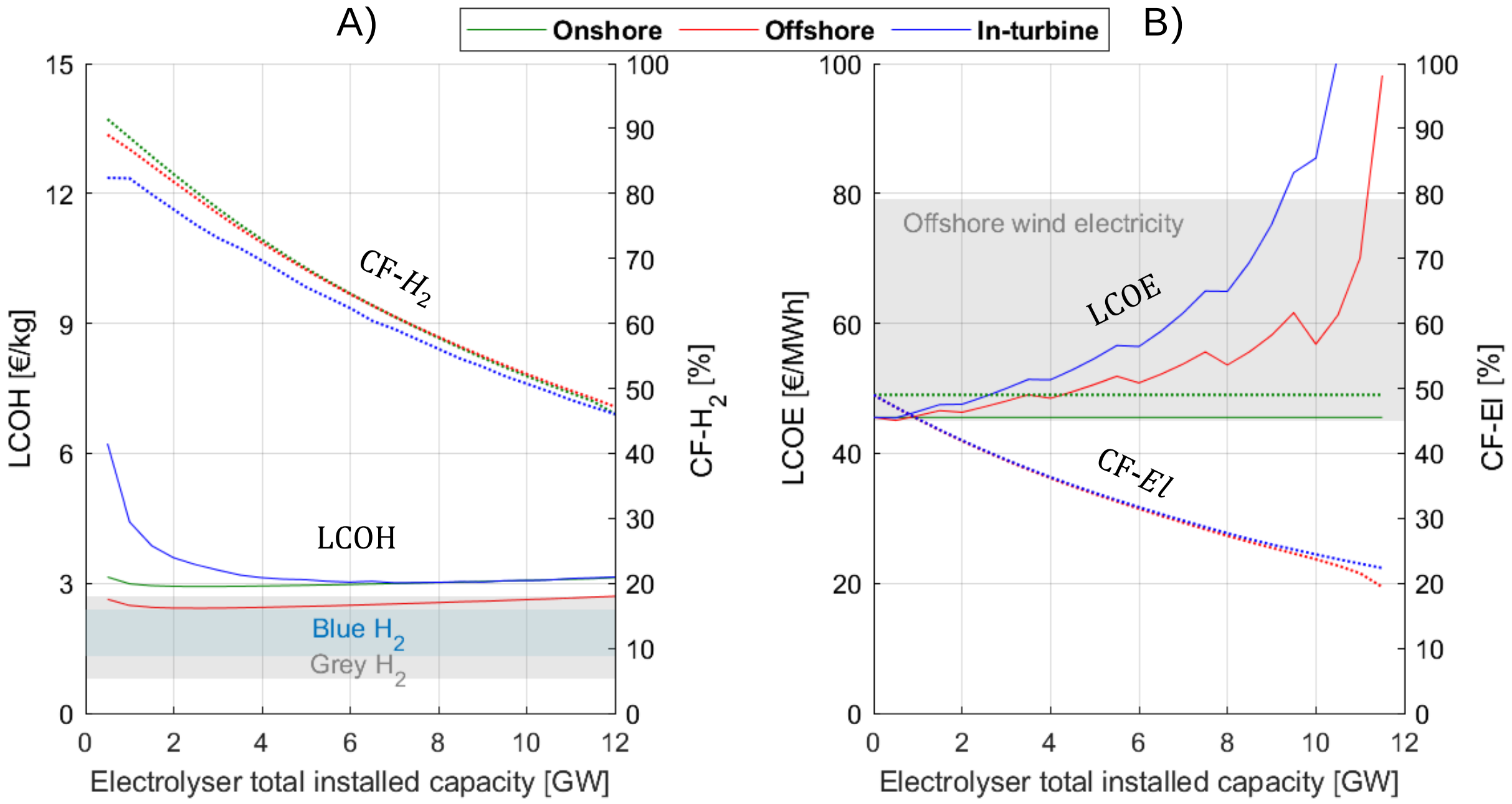

**Figure 6.** LCOH, LCOE and CF for the hydrogen-driven operation. CF- is the capacity factor of the electricity infrastructure (i.e. HVDC transmission cable, HVDC converters, substations, etc.) from the Hub to shore.

### 3.3 Electricity-driven operation mode

**Figure 7** presents the LCOH and LCOE for the electricity-driven mode of operation. As the capacity of the electricity infrastructure is reduced, and – similar to the hydrogen-driven operation – the installed electrolyser capacity is increased by an equal amount, two effects are detected.

First, the utilisation of the offshore electricity infrastructure will always be higher than that of hydrogen (CF-$El$> CF-$H_2$). The lower the electricity infrastructure capacity is (illustrated by a larger electrolyser capacity in **Figure 7.B**), the higher the CF-$El$ is, and, consequentially the, lower the LCOE will be. The minimum LCOE across all electrolyser placements is 39.4 €/MWh, achieved by offshore electrolysis. Comparing this with the case in which no electrolyser is installed highlights the fact that offshore electrolysis used for peak shaving leads to a 13% reduction. Second, an increase of the electrolyser capacity increases also its utilisation (CF-$H_2$ in **Figure 7.A**). Therefore, while in the hydrogen-driven operation economies of scale were counteracting the drop in hydrogen infrastructure utilisation, here, inversely, the economies of scale and CF of the electrolyser co-act. As a result, the larger the installed capacity, the lower the LCOH. The lowest LCOH, equal to 2.7 €/kg, is found in the case in which all the generated electricity is used for hydrogen production.

It is important to remind that in the electricity-driven operation the priority is to cover the electricity demand, therefore the electricity losses of the electrical infrastructure upstream of the electrolyser affect the utilisation of the electrolyser. This is the reason why small capacities of onshore electrolysers are producing no hydrogen when the electricity demand is 11.5 GW or higher (i.e. electrolyser capacity of 500 MW).

A final remark about **Figure 7.B** relates to the LCOE. Beyond electrolyser capacities of 8-10 GW, the LCOE starts increasing dramatically, due to the reduced amount of electricity transported compared with the fixed costs of the offshore electricity infrastructure. Therefore, if more than 85% of the offshore wind power is directed towards hydrogen production, it might be preferable to have a full-hydrogen offshore Hub.

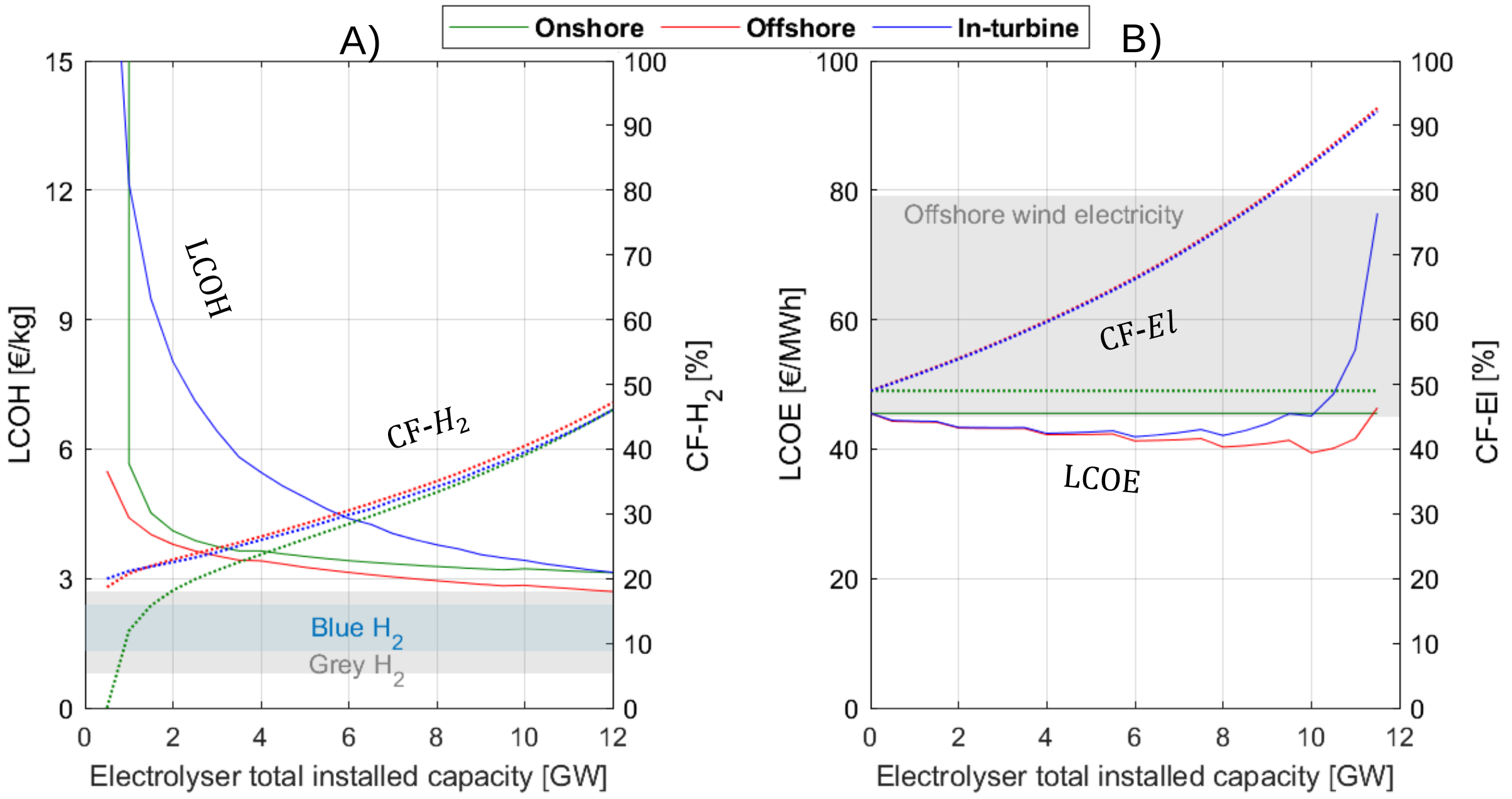

**Figure 7.** LCOH, LCOE and CF for the electricity-driven operation. CF-El is the capacity factor of the electricity infrastructure (i.e. HVDC transmission cable, HVDC converters, substations, etc.) from the Hub to shore.

## 3.4 Sensitivity analysis

### 3.4.1 Impact of cost of the components

In this section, the impact of the cost of each component on the median LCOH is assessed. This sensitivity analysis aims to indicate the effects on the LCOH of uncertainties that might affect the costs of each component. This is performed by individually changing ±25% the cost of each component, intending to determine which components have a larger impact on the LCOH. This shall provide insights about which components show the greatest potential for larger hydrogen cost reduction.

As shown in **Figure 8**, the cost of the WTs has a major impact, irrespective of the electrolyser placement and operation mode. For hydrogen-driven onshore electrolysis, the HVDC transmission is the second most relevant component. These results show that the cost of the electrical equipment upstream of the electrolyser is a major component of the LCOH.

Among the hydrogen infrastructure components, the cost of the electrolyser affects the most the median LCOH; this is especially noticeable in the in-turbine placement, where the cost of the electrolyser is penalised by small scales.

Moreover, for the in-turbine placement, both the pipeline and the compressor costs have a more significant impact compared to the other placements. This happens because, first, several small pipelines need to be installed to transfer the produced hydrogen from the OWPPs to the Hub, and second, due to pressure losses in these pipelines, larger compressor capacities are needed on the Hub when compared to the offshore and onshore placements.

It is also very interesting to observe that the costs of the desalination unit and the artificial island have a negligible effect on the LCOH, both in the hydrogen-driven and the electricity-driven operation.

In the case of the electricity-driven operation, the impact of the electrolyser is larger compared to the hydrogen-driven mode. This is because, at a parity of installed capacity (and CapEx), having a lower CF compared to the hydrogen-driven operation, the electrolyser produces less hydrogen, thus resulting in a higher cost per unit of kilogram of hydrogen delivered. Therefore, a change in the cost of the electrolyser, and the hydrogen pipelines, affects more heavily the LCOH.

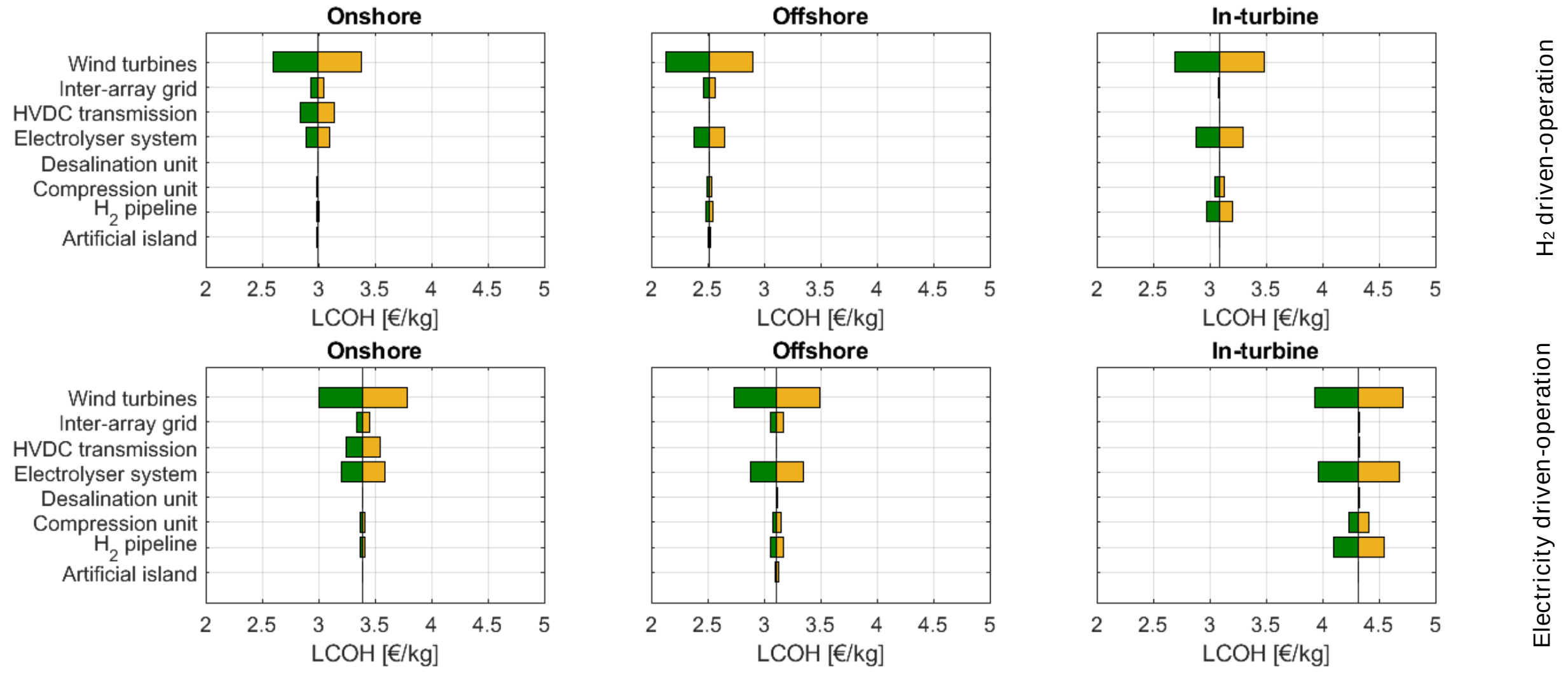

**Figure 8**. Median LCOH obtained by a perturbation of ±25% of the CapEx of each component. Median LCOH: the median of the LCOHs calculated considering 0.5 GW to 12 GW electrolyser installed capacities

### 3.4.2 Impact of the distance from shore

Since the exact location of the first Hub is still uncertain and several Hubs are expected to be constructed in the North Sea, in this section, the impact of the distance of the Hub from shore on the median LCOH was investigated. This analysis is, therefore, limited to all the components depending on the variable $L$ in this study.

As shown in **Figure 9**, there is a slight increase of the median LCOH with increasing distances, but the changes are mild. Offshore electrolysis maintains the lowest median LCOH across the range of possible distances, while in-turbine electrolysis maintains the highest.

It can be observed that the onshore electrolysis curve is steeper than the other two curves; this happens because, for onshore electrolysis, the HVDC transmission plays a major role in determining the LCOH (**Figure 8**). It is possible to conclude that the cost of HVDC lines is more sensitive to distance compared with the hydrogen pipelines, used for offshore and in-turbine electrolysis. Therefore, as the LCOH for in-turbine and offshore electrolysis is only marginally affected by the distance from shore, even more distant applications (i.e. far-offshore) would be possible.

It is also interesting to observe that in the case of hydrogen-driven operation, for shorter distances also the hydrogen produced with onshore electrolysis is competitive with grey hydrogen.

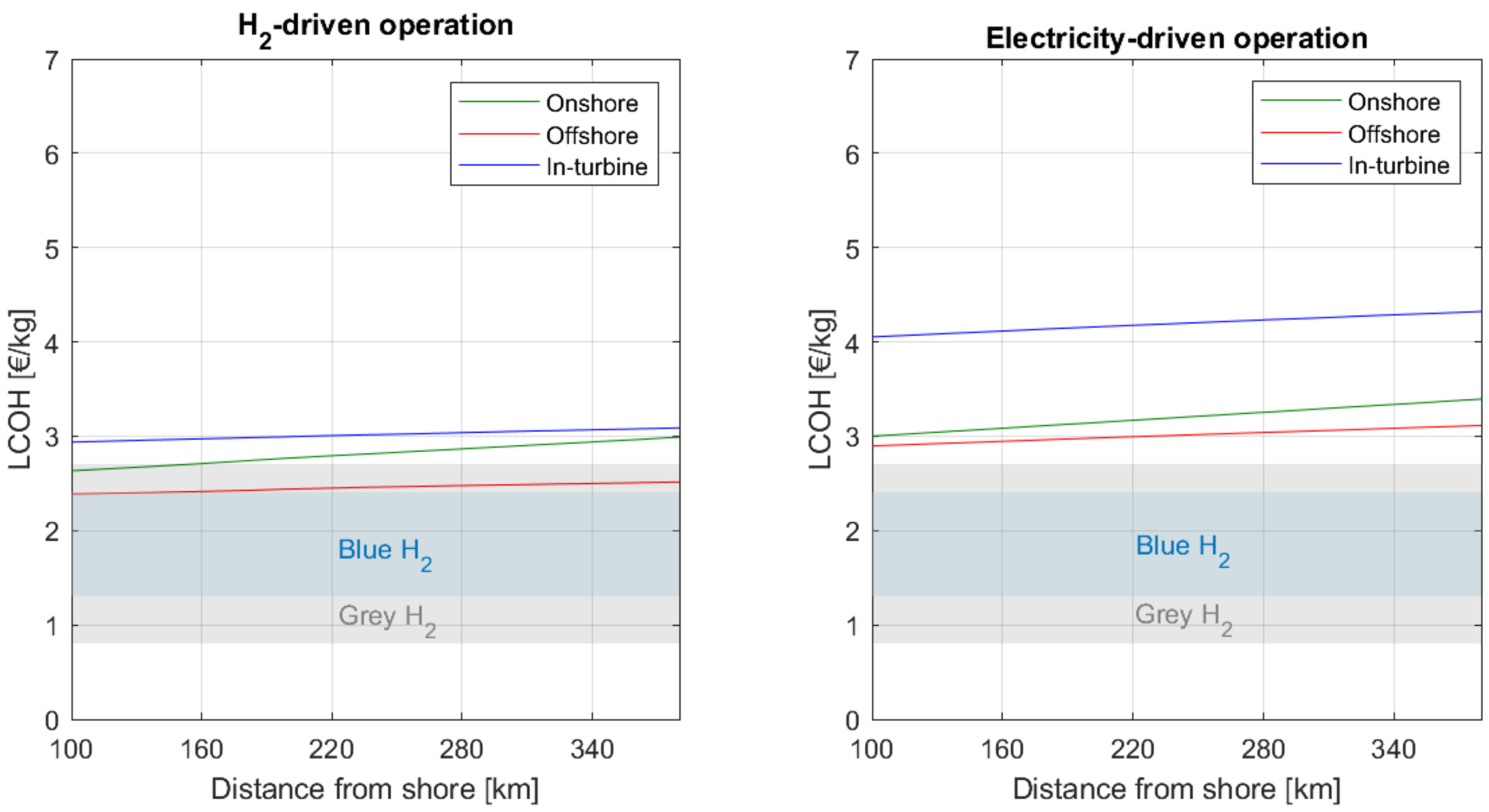

**Figure 9**. Median LCOH per distance of the Hub to shore.

## 4 Conclusions

The holistic techno-economic assessment proposed in this study assessed the cost of production of hydrogen and electricity from offshore wind power in the North Sea, comparing three different electrolyser placements (in-turbine, offshore and onshore), three technologies (alkaline, proton-exchange membrane and solid-oxide electrolysers), and two modes of electrolyser operation (hydrogen-driven and electricity-driven). Results showed that the different types of electrolysers are equally competitive, with the alkaline electrolyser achieving marginally lower costs. In terms of electrolyser placement, offshore electrolysis resulted in the lowest cost of hydrogen. The minimum LCOH, obtained for offshore electrolysis and hydrogen-driven operation mode, was estimated at 2.4 €/kg, which is competitive with the current costs of grey and blue hydrogen.

In the case of the electricity-driven operation of the electrolyser, the cost of electricity reduced up to 13% when compared to the LCOE without any electrolyser installed.

Offshore electrolysis is still not mature in terms of required infrastructure and integration with the offshore power systems, in particular for GW-scale electrolysis. Therefore, the input values for the hydrogen infrastructure have to be considered as estimations determined after discussion with manufacturers and operators. Only the major components of the systems are considered, to limit the complexity of the model and to generate results that would drive more in-depth studies. Moreover, social and environmental analyses were out of the scope. However, these aspects are also necessary to evaluate the feasibility of the placement.

Taking the presented results as a starting point, the cost of offshore green hydrogen can further reduce if the hydrogen infrastructure is more tightly integrated with:

- existing oil and gas infrastructures: e.g. repurposing platforms and pipelines.
- offshore electricity infrastructures: e.g. combining the transmission of hydrogen and electricity in a single component, instead of having separate cables and pipelines, and, thus, avoiding double installation costs; providing services to the electrical grid, such as a flexible resource for grid balancing.
- energy/industrial systems: e.g. benefitting from the synergies obtained by using by-products of electrolysis, such as oxygen and heat, and/or further converting hydrogen into e-fuels.

**Acknowledgements**

This research has been supported by the North Sea Pre-Feasibility Study project, funded by Det Energiteknologiske Udviklings- og Demonstrations Program (EUDP) under Grant.nr. 64018-058. The authors also would like to thank the Advisory Board Meeting members: Cenergy Holdings/Hellenic Cables, Dansk Energi, Dansk Industri, Energinet, Green Hydrogen Systems, Hitachi ABB, Ørsted, Semco Maritime, Siemens, Siemens-Gamesa Renewable Energy, Vestas, Wind Denmark.

**Appendix A**

The compressibility factor $Z$ is calculated using Eq. (A.1).

$$Z_{MEAN} = \frac{\left(\frac{p_{R,IN}}{p_{PIPE,IN}} + \frac{p_{R,OUT}}{p_{PIPE,OUT}}\right)}{2} \tag{A.1}$$

where $p_R$ is the pressure of the real gas, in kilopascal, calculated using Eq. (A.2), considering the inlet and outlet pressures of the pipelines.

$$p_R = \frac{R \cdot T_{MEAN}}{v(T_{MEAN}, p) - b} - \frac{a}{\sqrt[2]{T_{MEAN}} \cdot v(T_{MEAN}, p) \cdot (v(T_{MEAN}, p) + b)} \tag{A.2}$$

where $R$ is the universal constant of gas, 8.31434 J $mol^{-1}K^{-1}$; $v$ is the molar volume of the hydrogen in units of cubic metres per kilomole; and $a$ and $b$: factors of the Redlich-Kwong equations, calculated using Eq. (A.3) and Eq. (A.4).

$$a = 0.42748 \cdot \frac{R^2 \cdot T_C^{\frac{5}{2}}}{p_C} \tag{A.3}$$

$$b = 0.08664 \cdot \frac{R \cdot T_C}{p_C} \tag{A.4}$$

where $T_C$ is the critical temperature of hydrogen equivalent to 33.2 K; $p_C$ is the critical pressure of hydrogen equivalent to 1,320 kPa.

The coefficient of friction factor, or Darcy-Weisbach, $\lambda$ is calculated by solving the Colebrook-White equation for gas in pipelines in turbulent flows ($Re > 4000$), Eq. (A.5).

$$\frac{1}{\sqrt[2]{\lambda}} = -2 \cdot \log\left(\frac{K}{3.7} + \frac{2.51}{Re \cdot \sqrt[2]{\lambda}}\right) \tag{A.5}$$

where $K$ is the roughness factor in a pipeline, calculated using Eq. (A.6).

$$K = \frac{\epsilon}{D} \tag{A.6}$$

where ε is the equivalent sand roughness, assumed 0.05 mm [38]; and $Re$ is the Reynolds number for the flow in a pipe used for gas pipeline design [35], calculated using Eq. (A.7).

$$Re = 0.5134 \cdot \left(\frac{P_b}{T_b}\right)\left(\frac{G_{HYD} \cdot \dot{V}_{H_2,PIPE}(T_b, p_b) \cdot 24}{\mu(T_b, p_b) \cdot D}\right) \tag{A.7}$$

where $\mu$ is the dynamic viscosity of hydrogen at standard conditions, $8.64\ 10^{-5}$ poise.

The erosional velocity is calculated, $u_{MAX}$ in units of metres per second [35], using Eq. (A.8).

$$u_{MAX} = 100\sqrt{\frac{Z \cdot R \cdot T_{MEAN}}{29 \cdot G_{HYD} \cdot P}} \tag{A.8}$$

Acceptable operational velocity, $u$, is assumed to be lower than 50% of the erosional velocity in units of metres per second [35].

Having a known mass flow rate, the velocity $u$ is related to the diameter of the pipeline $D$, according to Eq. (A.9).

$$\dot{m}_{H_2}(t) = u \cdot \rho(T, p) \cdot \pi \cdot \frac{D^2}{4} \tag{A.9}$$

where ρ is the density in units of kilograms per cubic metre.

For the pipeline from the Hub to the shore, the diameter is calculated using Eq. (A.9), with $p = p_{TRANS}$, and $\rho(T_{MEAN}, p_{TRANS})$. $p_{PIPE,IN}$ is then calculated solving Eq. (22). $p_{PIPE,IN}$ is then equivalent to $p_{COMP,OUT}$ and it can be used in Eq. (21).

For the pipelines from the OWPPs to the Hub, diameter, *D*, and outlet pressure, $p_{PIPE,OUT}$, are found maximising $u_{PIPE,OUT}$, considering two constraints: $u_{PIPE,OUT} < 0.5 \cdot u_{MAX}$ and $p_b \leq p_{PIPE,OUT} \leq p_{PIPE,IN}$. The higher the velocity the lower is the diameter, having a fixed mass flow rate, thus reducing the cost.

# Appendix B

**Table B.1**. Cost inventory for the calculation of LCOE and LCOH.

| Component | Symbol | Value | Comment | Ref. |
|---|---|---|---|---|
| **Capital expenditures, CapEx [M€]** | | | | |
| Wind power plant | $CapEx_{OWPP,EQ}$ | $(14+7.55)\cdot N_{WT}$ | 14 M€ represents the cost of all components of a reference 15 MW; 7.55 M€ represents the costs of the substructure underneath the turbine, determined based on the required mass of the tower, transition piece and monopile foundation for an average depth of 30 m. | [20] |
| | $CapEx_{OWPP,NEQ}$ | $100\cdot P_{OWPP}$ | This equation represents the project development, including all costs up to the start of construction. | [20] |
| Inter-array grid | $CapEx_{IG,EQ}$ | $\sum_i L_{IG,i}\cdot\left[94.94\cdot 10^{-3}\ +86.2\cdot 10^{-3}\ \cdot e^{\left(\frac{2.05\cdot P_{IG}}{10^8}\right)}\right]$ | Cost equation of AC cables based on a rated voltage of 66 kV. | [39] |
| | $CapEx_{IG,NEQ}$ | $\sum_i 0.33\ \cdot L_{IG,i}$ | Installation costs for offshore inter-array grid cables. | [39] |
| Offshore substation | $CapEx_{SS,OFF}$ | $117.9\cdot P_{HVDC}+\left\lceil\frac{P_{HVDC}}{P_{MAX}}\right\rceil\cdot 45.4$ | Curve fitting the average of the cost of an offshore VSC-HVCDC. $P_{MAX}$ is considered 2 GW. | [40] |
| Onshore substation | $CapEx_{SS,ON}$ | $101\cdot P_{HVDC}+\left\lceil\frac{P_{HVDC}}{P_{MAX}}\right\rceil\cdot 61.6$ | Curve fitting the average of the costs onshore VSC-HVDC substations. $P_{MAX}$ is considered 2 GW. | [40] |
| VSC-HVDC transmission | $CapEx_{HVDC}$ | $L_{HS}\cdot\left[0.6\cdot P_{HVDC}+\left\lceil\frac{P_{HVDC}}{P_{MAX}}\right\rceil\cdot 1.345\right]$ | Curve fitting including HVDC extruded copper 320-400 kV and the installation and the average of 2 single cables; 2 trenches, single-core, 10m apart. $P_{MAX}$ is considered 2 GW. | [40] |
| Electrolyser system | $CapEx_{EL}$ | $P_{ELEC}\cdot RC_{ELEC}\cdot(1+IF\cdot OS)\cdot\left(\frac{P_{ELEC}\cdot 10^3}{RP_{ELEC}}\right)^{SF_{ELEC}}$ | Non-equipment costs: land, contingency, contractors, legal fees, construction, engineering, yard improvements, buildings, electrics, piping, instrumentation and installation and grid connection. The cost for the offshore configuration is assumed to be double the onshore costs. (OS =1 if the electrolyser is located in-turbine or offshore, OS =0 if the electrolyser is located onshore, reflecting Siemens estimations). | [34,41] |
| Desalination unit | $CapEx_{DES}$ | $30.6\cdot\bar{V}_{DES}$ | Reverse osmosis seawater desalinator (Lenntech Reverse Osmosis System) is used as reference technology. | [16] |
| Compression unit | $CapEx_{COMP}$ | $3{,}000\cdot P_{COMP}$ | Considering a centrifugal compressor with electric drivers, including power lines, transformers, and electronics. | [42] |
| Hydrogen pipeline | $CapEx_{PIPE}$ | $1.75\cdot L_{HS}\cdot[0.314+0.574\cdot 10^3\cdot(D)+1.7\cdot 10^6\cdot(D)^2]$ | Pipeline for hydrogen transmission in the North Sea. | [16] |
| Artificial island | $CapEx_{HUB}$ | $(3.26\ \cdot V_{HUB}+804\cdot A_{HUB})\cdot 10^{-6}$ | The cost of dredged sand is assumed to be 3.26 €/$m^3$ and the cost for protecting the shoreline of the island is assumed to be 804 €/ $m^2$. The cost of the artificial island is assumed to be allocated to the electricity and hydrogen generated proportionally to the footprint of their components, HVDC offshore substation for the electricity system, and electrolyser for the hydrogen | [43] |
| **Operation and maintenance expenditures, OpEx [M€/a]** | | | | |
| Wind power plant | $OpEx_{OWPP}$ | $1.9\% \cdot CapEx_{OWPP,EQ}$ | - | [26] |
| Inter-array grid | $OpEx_{\mathrm{IG}}$ | $0.2\% \cdot CapEx_{IG,EQ}$ | - | [44] |
| VSC-HVDC transmission | $OpEx_{HVDC}$ | $0.2\% \cdot CapEx_{HVDC}$ | $CapEx_{HVDC}$ includes the cost of the substations and the transmission line. | [44] |
| Electrolyser system | $OpEx_{ELEC,EQ}$ | $CapEx_{ELEC} \cdot \left(1 - IF \cdot (1 + OS)\right) \cdot 3.44\% \cdot (P_{ELEC} \cdot 10^3)^{-0.155}$ | Including material cost for planned and unplanned maintenance, labour cost in central Europe, which all depend on a system scale. Excluding the cost of electricity and the stack replacement, calculated separately. Scaled maximum to $\bar{P}_{ELEC}$ = 1 GW. | [32] |
| | $OpEx_{ELEC,SR}$ | $P_{ELEC} \cdot RC_{SR} \cdot \left(\frac{P_{ELEC} \cdot 10^3}{RP_{SR}}\right)^{SF_{SR}} \cdot \left\lfloor \frac{OH}{OH_{MAX}} \right\rfloor$<br>$RC_{SR} = RU_{SR} \cdot RC_{ELEC} \cdot (1 - IF) \cdot \left(\frac{RP_{SR}}{RP_{ELEC}}\right)^{SF_{ELEC}}$<br>$SF_{SR} = 1 - (1 - SF_{SR,0}) \cdot e^{-\frac{\bar{P}_{ELEC}}{\bar{P}_{STACK,MAX}}}$ | Approximation of stack costs and replacement cost depending on the electrolyser equipment costs. Paid only the year in which the replacement is needed. | [34,45] |
| | $OpEx_{ELEC,NEQ}$ | $4\% \cdot CapEx_{ELEC} \cdot IF \cdot (1 + OS)$ | It covers the other operational expenditure related to the facility level. This includes site management, land rent and taxes, administrative fees (insurance, legal fees…), site maintenance. | [34] |
| Desalination unit | $OpEx_{DES}$ | $2.5\% \cdot CapEx_{DES}$ | Operational expenditure of desalination when assumed part of the electrolyser system. | [16] |
| Compression unit | $OpEx_{COMP}$ | $4\% \cdot CapEx_{COMP}$ | Fixed operational and maintenance costs. | [46] |
| Hydrogen pipeline | $OpEx_{PIPE}$ | $2\% \cdot CapEx_{PIPE}$ | Fixed operational and maintenance costs for both $CapEx_{PIPE,HS}$ and $CapEx_{PIPE,WTH}$. | [16] |
| Freshwater | $OpEx_{H_2O}$ | $9.23 \cdot (1 - 0.6) \cdot 10^{-6} \cdot \sum_{t=1}^{8760} \dot{V}_{H_2O,DES}(t)$ | In the case of offshore electrolysis, water is purchased from the grid. 9.23 € per cubic meter of water is assumed as an average price and a 60% discount for large consumers. | [47] |

Conversions used from the original currencies: $USD_{2014}$=0.752 $EUR_{2014}$, EUR inflation from 2014 to 2017 = 1.81%; EUR inflation from 2010 to 2017 = 9.11%; $SEK_{2003}$ to = 0.1096 $EUR_{2003}$, EUR inflation from 2003 to 2017 = 25.33%; $GBP_{2015}$ = 1.35 $EUR_{2015}$, EURO inflation from 2015 to 2017 = 1.78%.

**Electrolyser economies of scale**

Large scale electrolysers are still under development, so no commercial cost reference exists. However, an investigation conducted by Zauner et al.[33] showed that the effect of economies of scale is more pronounced at lower nominal power levels than at higher levels. This leads to an increased share of stack costs in the overall system for larger electrolysis systems, which reduces the overall effect of the economies of scale. In this study, it is assumed that the scale factor for small units is used to calculate the costs for electrolysers not larger than 10 MW, while the scale factor for large sizes is used for electrolysers larger than 10 MW. It is also assumed that no additional economies of scale are accounted for in sizes larger than 100 MW. The average costs for the different technologies for 2030 has been sourced from the Energinet Technology Catalogue[30]. (**Table 4**).

**Table 4**. Coefficients used for $CapEx_{EL}$ calculations (sourced from[30,33]).

| | **Reference cost, $RC_{ELEC}$ [€/kW]** | **Installation fraction*, IF [%$RC_{ELEC}$]** | **Reference power, $RP_{ELEC}$ [MW]** | **Scale factor, $SF_{ELEC}$ [<10 MW/>10 MW]** |
|---|---|---|---|---|
| AEL | 550 | 27 | 10 | -0.24/-0.13 |
| PEMEL | 600 | 33 | 10 | -0.21/-0.14 |
| SOEL | 600 | 63 | 15 | -0.25/-0.22 |

*Installation costs include: land, contingency, contractors, legal fees, construction, engineering, yard improvements, buildings, electrics, piping, instrumentation and installation and grid connection

The economies of scale of each piece of the equipment composing the electrolyser system (i.e. stack, power electronics, gas conditioning, gas conditioning, balance of plant) are different. Therefore, the cost of the stack would not follow the economies of the entire electrolyser unit. The stack does not show potential for large cost reduction via economies of because of its modular design [33]. The values used in the calculations are listed in **Table 5**.

**Table 5**. Coefficients used for $OpEx_{EL,SR}$ calculations (sourced from[33]).

| | Reference cost share*, $RU_{SR}$ [%] | Average max size, $P_{STACK,MAX}$ [MW] | Average scale factor, $SF_{SR,0}$ |
|---|---|---|---|
| AEL | 45 | 4 | 0.12 |
| PEMEL | 41 | 2 | 0.11 |
| SOEL | 50 | 1 | 0.13 |

*for a reference power, $RP_{SR}$, of 5 MW.

## References

[1] European Commission. The European Green Deal. Brussels, Belgium: 2019. doi:10.1017/CBO9781107415324.004.

[2] Folketinget. Klimaaftale for energi og industri mv. 2020. Copenhagen K, Denmark: 2020.

[3] IRENA. Hydrogen : a Renewable Energy Perspective. Abu Dhabi: 2019.

[4] Energistyrelsen. Cost benefit analyse og klimaaftryk af energiøer i Nordsøen og Østersøen Cost benefit analyse og klimaaftryk af energiøer i Nordsøen og Østersøen. 2021.

[5] North Sea Wind Power Hub Consortium. Concept Paper 4: Towards Spatial Planning of North Sea Offshore Wind. 2019.

[6] Ørsted. A European Green Deal - How offshore wind can help decarbonise Europe. 2019.

[7] The European Parliament and the Council of the European Union. Offshore Wind Energy in Europe. 2020.

[8] European Commission. An EU Strategy to harness the potential of offshore renewable energy for a climate neutral future. Brussel, Belgium: 2020.

[9] North Sea Wind Power Hub Consortium. Modular Hub-and-Spoke Concept to Facilitate

Large Scale Offshore Wind. 2019.

[10] Weichenhain U, Elsen S, Zorn T, Kern S. Hybrid projects : How to reduce costs and space of offshore developments North Seas Offshore Energy Clusters study. 2019.

[11] North Sea Wind Power Hub Consortium. Concept Paper 3: Modular Hub-and-Spoke: Specific solution options. 2019.

[12] Energinet. Winds of Change In A Hydrogen Perspective - PtX Strategic Action Plan. 2019.

[13] European Commission. The hydrogen strategy for a climate-neutral Europe. 2020.

[14] Meier K. Hydrogen production with sea water electrolysis using Norwegian offshore wind energy potentials: Techno-economic assessment for an offshore-based hydrogen production approach with state-of-the-art technology. Int J Energy Environ Eng 2014;5:1–12. doi:10.1007/s40095-014-0104-6.

[15] Jepma C, Van Schot M. On the economics of offshore energy conversion: smart combinations_Converting offshore wind energy into green hydrogen on existing oil and gas platforms in the North Sea. 2017.

[16] Jepma C, Kok G-J, Renz M, van Schot M, Wouters K. North Sea Energy D3.6 Towards sustainable energy production on the North Sea-Green hydrogen production and CO2 storage: onshore or offshore? As Part of Topsector Energy: TKI Offshore Wind & TKI New Gas: 2018.

[17] Crivellari A, Cozzani V. Offshore renewable energy exploitation strategies in remote areas by power-to-gas and power-to-liquid conversion. Int J Hydrogen Energy 2020;45:2936–53. doi:10.1016/j.ijhydene.2019.11.215.

[18] International Energy Agency. Hydrogen production costs by production source 2020. https://www.iea.org/data-and-statistics/charts/hydrogen-production-costs-by-production-source-2018 (accessed March 14, 2021).

[19] European Commission. Report from the commission to the European Parliament and the Council on progress of clean energy competitiveness. 2020.

[20] E.C.M. Ruijgrok PhD, E.J. van Druten MSc BHBMs. Cost Evaluation of North Sea Offshore Wind Post 2030. Petten, The Netherlands: 2019. doi:112522/19-001.830 112522.

[21] Swamy SK, Saraswati N, Warnaar P. North Sea Wind Power Hub ( NSWPH ): Benefit study for ( 1 + 3 ) potential locations of an offshore hub- island. Petten (The Netherlands): 2019. doi:06.37770.

[22] The MathWorks Inc. MATLAB and Statistics Toolbox Release 2019b. Natick, Massachusetts, United States: 2016.

[23] Goodwin DG, Moffat HK, Speth RL. Cantera: An object- oriented software toolkit for chemical kinetics, thermodynamics, and transport processes. Pasadena, CA: Caltech: 2017.

[24] ECMWF. ERA5 hourly data on single levels from 1979 to present. 2018. doi:10.24381/cds.adbb2d47.

[25] Gaertner E, Rinker J, Sethuraman L, Anderson B, Zahle F, Barter G. IEA Wind TCP Task 37: Definition of the IEA 15 MW Offshore Reference Wind Turbine. United States: 2020. doi:doi:10.2172/1603478.

[26] The Danish Energy Agency, Energinet. Technology Data - Generation of Electricity and District heating 2016:414.

[27] Greedy, Lyndon. TENNET, NL OFFSHORE WIND FARM TRANSMISSION SYSTEMS 66 kV Systems for Offshore Wind Farms 2015:35.

[28] U.S. Department of Energy. Assessing HVDC Transmission for Impacts of Non - Dispatchable Generation 2018:1–32.

[29] IEA. The Future of Hydrogen - Seizing today's opportunities. Rep Prep by IEA G20, Japan 2019. doi:10.1787/1e0514c4-en.

[30] Danish Energy Agency and Energinet. Technology Data for Renewable Fuels - Technology descriptions and projections for long-term energy system planning (2020 updated). 2017.

[31] Schmidt O, Gambhir A, Staffell I, Hawkes A, Nelson J, Few S. Future cost and performance of water electrolysis: An expert elicitation study. Int J Hydrogen Energy 2017;42:30470–92. doi:10.1016/j.ijhydene.2017.10.045.

[32] Bertuccioli L, Chan A, Hard D, Lehner F, Madden B, Standen E. Development of water electrolysis in the European Union. vol. 23. 2014.

[33] Zauner A, Böhm H, Rosenfeld DC, Tichler R. Innovative large-scale energy storage technologies and Power-to-Gas concepts after optimization. D7.7 Analysis on future technology options and on techno-economic optimization 2019:1–89.

[34] Tractebel E, Engie, Hinicio. Study on Early Business Cases for H2 in Energy Storage and More Broadly Power To H2 Applications. EU Comm 2017:228.

[35] Menon ES. Pipeline planning and construction field manual. The Boulevard, Langford Lane, Kidlingron, Oxford, OX5 1GB, UK: Gulf Professional Publishing, Elsevier Inc.; 2011.

[36] Weber AC, Papageorgiou LG. Design of hydrogen transmission pipeline networks with hydraulics. Chem Eng Res Des 2018;131:266–78. doi:10.1016/j.cherd.2018.01.022.

[37] Electricity Ten Year Statement 2015. Appendix E: Electricity Ten Year Statement 2015:2015. doi:10.1016/B978-0-08-091906-5.00027-6.

[38] Renz M, Schot M Van, Jepma C. North Sea Energy Energy transport and energy carriers 2020.

[39] Lundberg S. Performance comparison of wind park configurations. Power Eng 2003.

[40] National Grid. Electricity Ten Year Statement 2015. UK Electr Transm 2015:1–145.

[41] Siemens, Personal Communication on electrolyser offshore installation cost. 2020.

[42] CEER. Pan-European cost-efficiency benchmark for gas transmission system operators 2019.

[43] Gerrits S, Kuiper C, Quist P, Van Druten EJ. Feasibility Study of the Hub and Spoke Concept in the North Sea: Developing a Site Selection Model to Determine the Optimal Location. Delft University of Technology, 2017.

[44] Das K, Antionios Cutululis N. Offshore Wind Power Plant Technology Catalogue - Components of wind power plants, AC collection systems and HVDC systems. Baltic Grid: 2017.

[45] IRENA. Hydrogen From Renewable Power: Technology outlook for the energy transition. 2018.

[46] Reuß M, Grube T, Robinius M, Preuster P, Wasserscheid P, Stolten D. Seasonal storage and alternative carriers: A flexible hydrogen supply chain model. Appl Energy

2017;200:290–302. doi:10.1016/j.apenergy.2017.05.050.

[47] DANVA. Water in figures. Godthåbsvej 83, DK-8660 Skanderborg: 2019.

**Supplementary material**

# Onshore, offshore or in-turbine electrolysis? Techno-economic overview of alternative integration designs for green hydrogen production into Offshore Wind Power Hubs

Alessandro Singlitico, Jacob Østergaard, Spyros Chatzivasileiadis

Center for Electric Power and Energy (CEE), Department of Electrical Engineering, Technical University of Denmark (DTU), 2800 Kgs. Lyngby, Denmark.

Corresponding author: alesi@elektro.dtu.dk

# 1 Input summary

| | | |
|---|---|---|
| **Thermodynamic values** | | |
| $LHV_{HYD}$ | 33.33 kWh/kg | |
| $T_{MEAN}$ | 285.15 K | |
| $T_{BASE}$ | 288.15 K | |
| $p_{BASE}$ | 101,325 Pa | |
| $G_{HYD}$ | 0.0696 [-] | |
| R | 8.31434 J/mol K | |
| $T_C$ | 33.2 K | |
| $p_c$ | 1,320 kPa | |
| $\mu$ | $8.64\ 10^{-5}$ poise | |
| **Technological values** | | |
| $P_{WT}$ | 15 MW | [1] |
| $N_{WT}$ | 5 | [2] |
| $SP_{WT}$ | 4.5 $MW/km^2$ | [3] |
| $\eta_{IG}$ | 0.55 % | [4] |
| $\eta_{HS}$ | 0.0035 % | [5] |
| $\eta_{ST}$ | 1 % | [5] |
| $W_{DES}$ | 15 l/kg | [6] |
| $e_{DES}$ | 3.5 $kWh/m^3$ | [7] |
| $\eta_{COMP}$ | 50% | [6] |
| ε | 0.05 mm | [8] |
| $f_{HVDC}$ | 4’860 $m^2$ /GW | [9] |
| h | 33 m | [2] |
| $L_{HS}$ | 380 km | [10] |
| $P_{HUB}$ | 12 GW | [2] |
| $P_{OWPP}$ | 1 GW | [2] |
| $p_{TRANS}$ | 70 bar | [11] |
| **Economic values** | | |
| DR | 5% | [11] |
| LT | 30 years | [2] |

## 2 Extended results

An initial overview of the specific CapEx of each unit involved in the $H_2$ infrastructure is shown in **Figure 1**. AEL presents specific CapEx per unit of capacity installed lower than PEMEL and SOEL.

The economies of scale of the electrolyser are visible in the in-turbine configuration, in which the electrolyser size ranges from 625 kW to 15 MW. In the case of offshore and onshore electrolysis, the cost per installed capacity of the electrolyser is constant, since over 100 MW the economies of scales are assumed not to affect the unitary cost.

The CapEx of the pipelines for distribution (i.e. from the OWPPS to the Hub) and transmission (i.e. from the Hub to shore) is affected by the number of the pipelines and their diameter. In the in-turbine case, because of the large number of pipelines involved and their small diameters, the CapEx of the distribution pipelines is a substantial share of the cost, especially in the case of small installed capacities, due to economies of scales of the pipeline. The effect of the economies of scales is evident also in the CapEx of the transmission pipeline in the case of offshore electrolysis.

The CapEx of the compressor is affected by its pressure ratio ($p_{OUT}/p_{IN}$): the larger the pressure ratio, the larger the CapEx of the compressor. The operating pressure of the electrolyser and the placement of the electrolyser affect the pressure ratio of the compressor. Higher operating pressure of the electrolyser results in a lower additional compression. Moreover, the closer to shore is the placement, the lower is the pressure ratio, since fewer are the pressure losses (i.e. in the pipelines from the OWPPs to the Hub and from the Hub to shore). The CapEx of the artificial island and the desalination unit have a minor share on the overall CapEx.

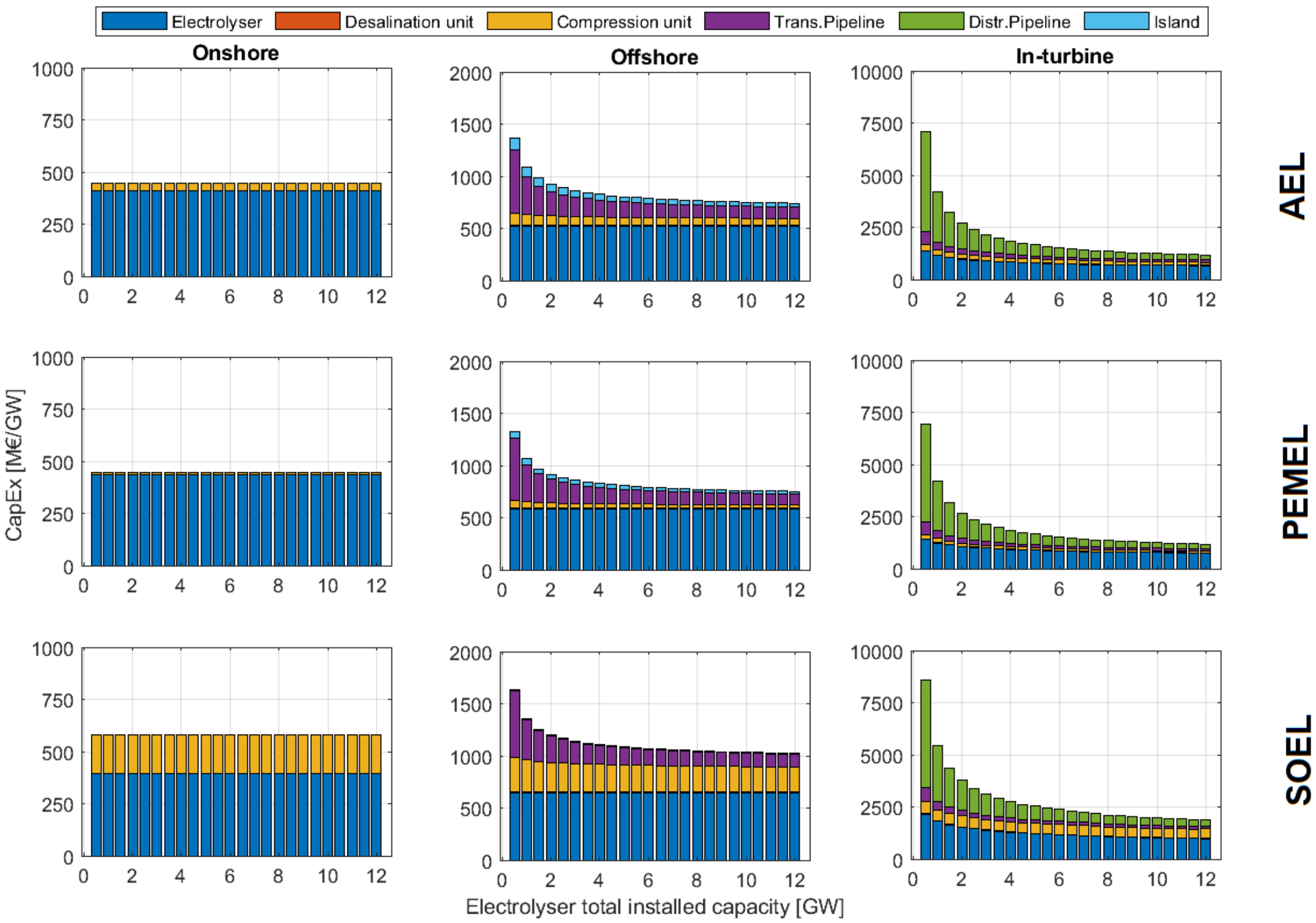

**Figure 1**. CapEx per unit of installed capacity. Note: these results are valid for both $H_2$-driven and electricity-driven operation of the electrolyser.

## 2.1 $H_2$-driven operation

OpEx per unit of installed capacity for the $H_2$ infrastructure is presented in **Figure 2**. The stack replacement is a major cost. The effect of the economies of scale is visible in the in-turbine case. Two stack replacements take place during the lifetime of the AEL and PEMEL electrolyser, and four stack replacements are required for the SOEL electrolyser, due to the lower amount of maximum operating hours. Purchasing freshwater to the onshore electrolyser is a major cost.

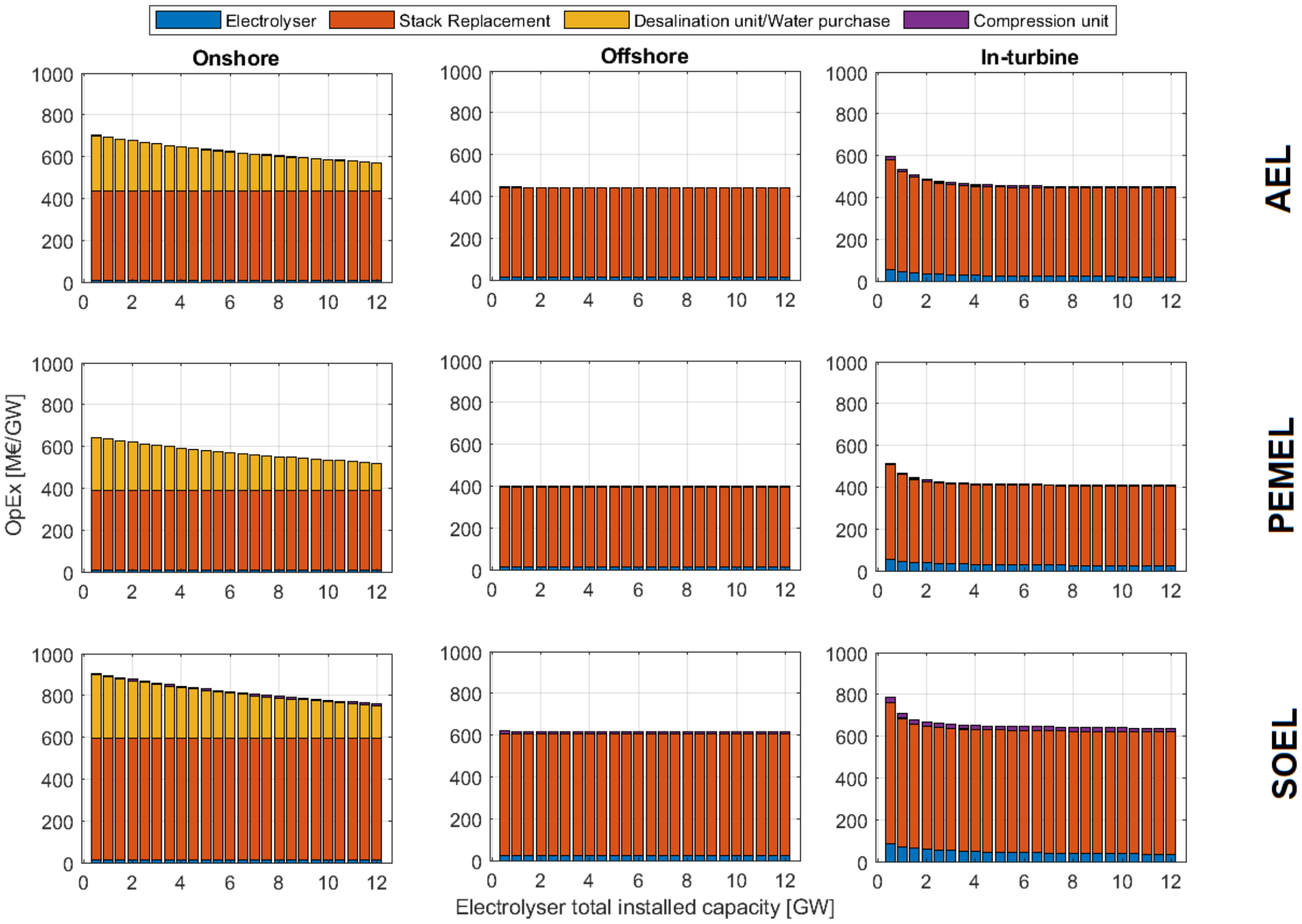

**Figure 2**. OpEx per unit of installed capacity, in the case of $H_2$-driven operation.

The energy consumption and the energy delivered in form of $H_2$ per unit of installed capacity is presented in **Figure 3**. Both energy consumption and delivered decrease by increasing the installed capacity, due to lower capacity factors. AEL and PEMEL show similar performance. SOEL, although a better nominal efficiency, is penalised by slower cold start-up and by the higher stack degradation. Therefore, for a consumption identical to AEL and PEMEL, the energy delivered is less. Moreover, the consumption of the compressor for SOEL is higher than for AEL and PEMEL, because of a higher pressure ratio, thus reducing the electricity directed to the electrolyser.

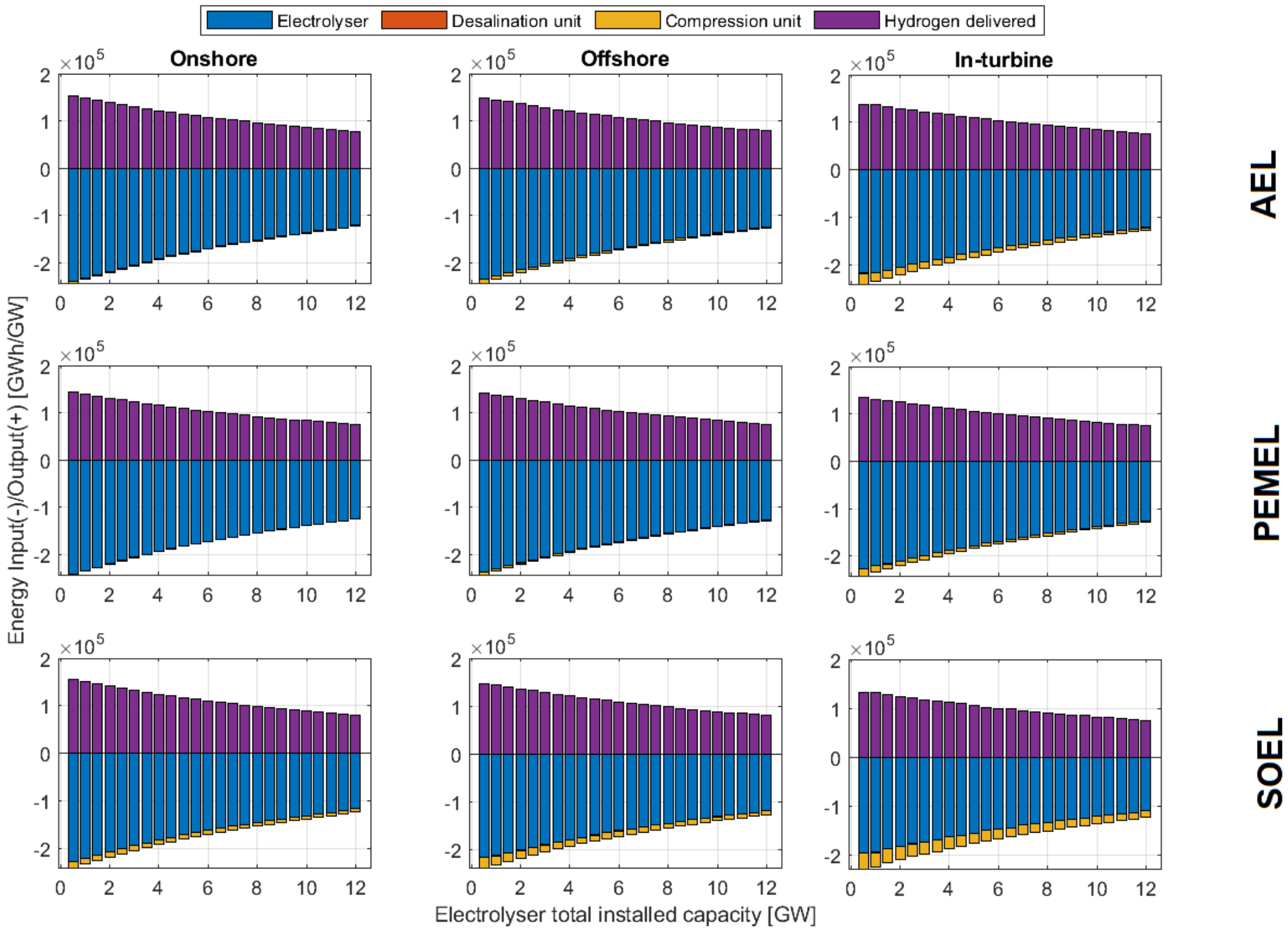

**Figure 3.** Energy consumed in form of electricity and delivered in form of $H_2$, in the case of $H_2$-driven operation.

The cost of purchasing electricity per unit of installed capacity is presented in **Figure 4**. This is an order of magnitude higher than the OpEx, resulting then to be the main cost driver, and also greater than the CapEx. The cost for the electricity purchased per unit of installed capacity decreases by increasing the installed capacity, due to the decreasing capacity factor. The cost for the purchased electricity is lower for the in-turbine placement, followed by the offshore and, finally, for the onshore placement. This is due to the cost of the electrical infrastructure upstream the electrolyser allocated in the cost of the electrical energy consumed by the $H_2$ infrastructure, larger for the onshore placement, followed by the offshore and in-turbine placements.

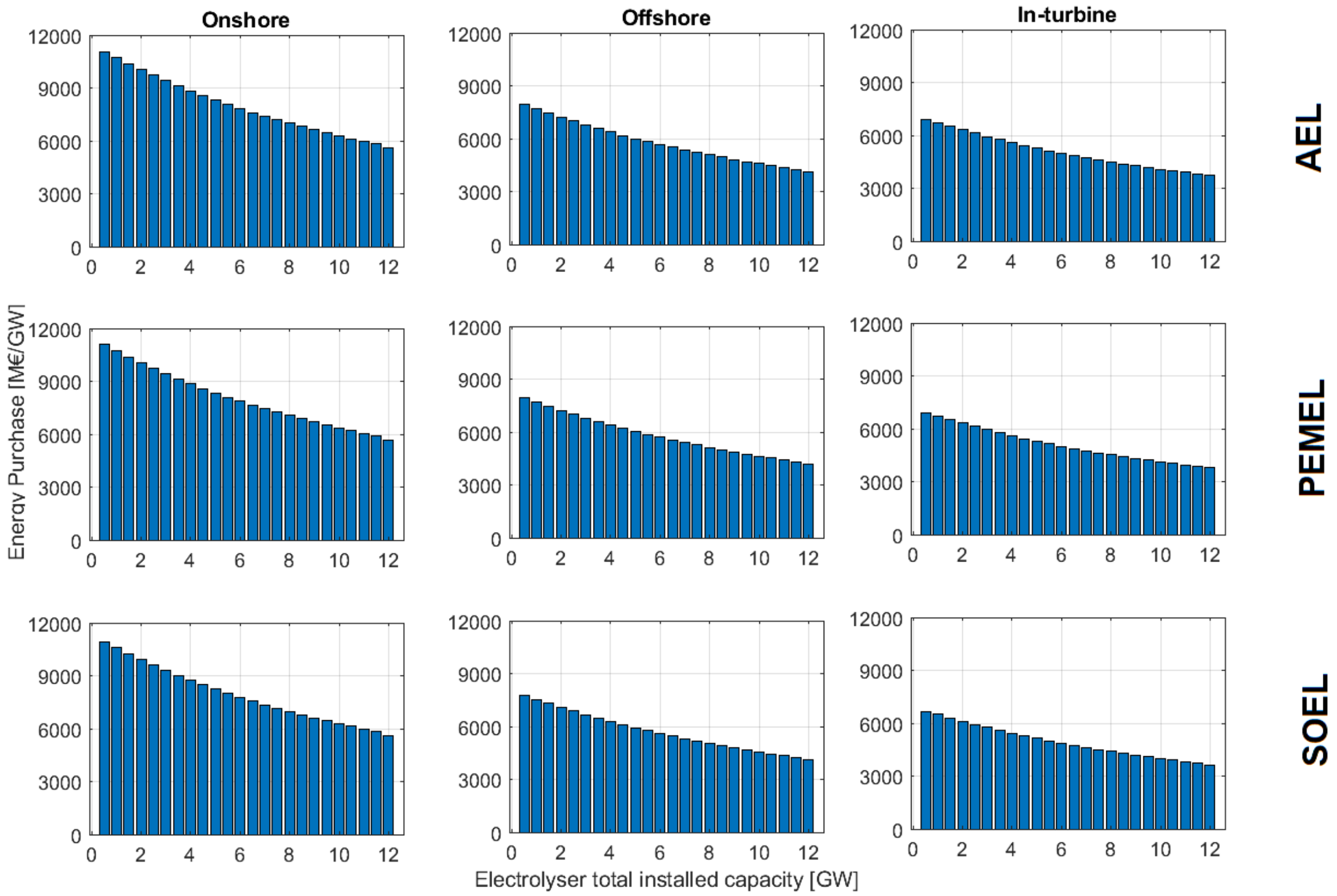

**Figure 4**. Electrical energy purchase per unit of installed capacity, in the case of $H_2$-driven operation.

The final results of LCOH and LCOE for each electrolyser technology and placement are presented in **Figure 5**. The LCOH is the results of the interrelated effects of CapEx, OpEx, and electrical energy purchase and energy delivered. Lower LCOHs are found for AEL, even though very close to PEMEL. Higher CFs are found for PEMEL, due to lower electricity diverted to the compression unit compared to AEL and SOEL, due to a higher operating pressure of the PEMEL. The low operating pressure of the SOEL, 5 bar, largely penalises this technology for in-turbine applications. LCOE and CF of the power transmission to shore are independent of the electrolyser technology used. Therefore, the discussion regarding the electricity infrastructure in the case of AEL (in the Article) can be extended to PEMEL and SOEL.

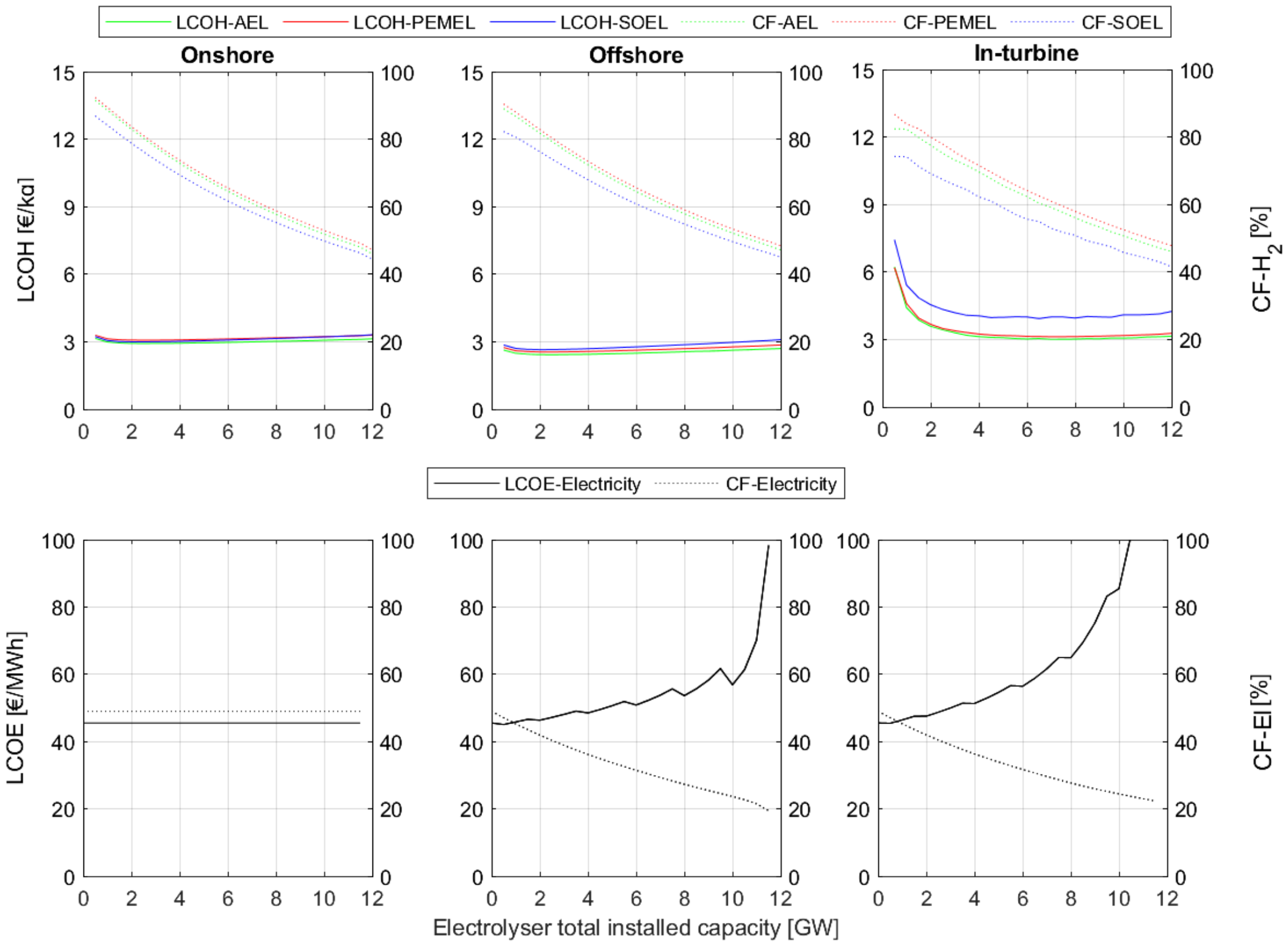

**Figure 5.** LCOH, LCOE and CF, in the case of $H_2$-driven operation. CF-electricity is the capacity factor of the electricity infrastructure (e.g. HVDC transmission cable, HVDC converters, substations, etc.) from the Hub to shore.

## 2.2 Electricity-driven operation

OpEx per unit of installed capacity for the $H_2$ infrastructure is presented in **Figure 6**. Differently from the $H_2$-driven operation, the stacks are not replaced for small sizes, due to operating hours lower than the maximum operating hours. The number of stack replacements increases with the electrolyser total installed capacity, due to the increasing CF (determining an increasing number of operational hours), up to two for AEL and PEMEL, and up to four for SOEL.

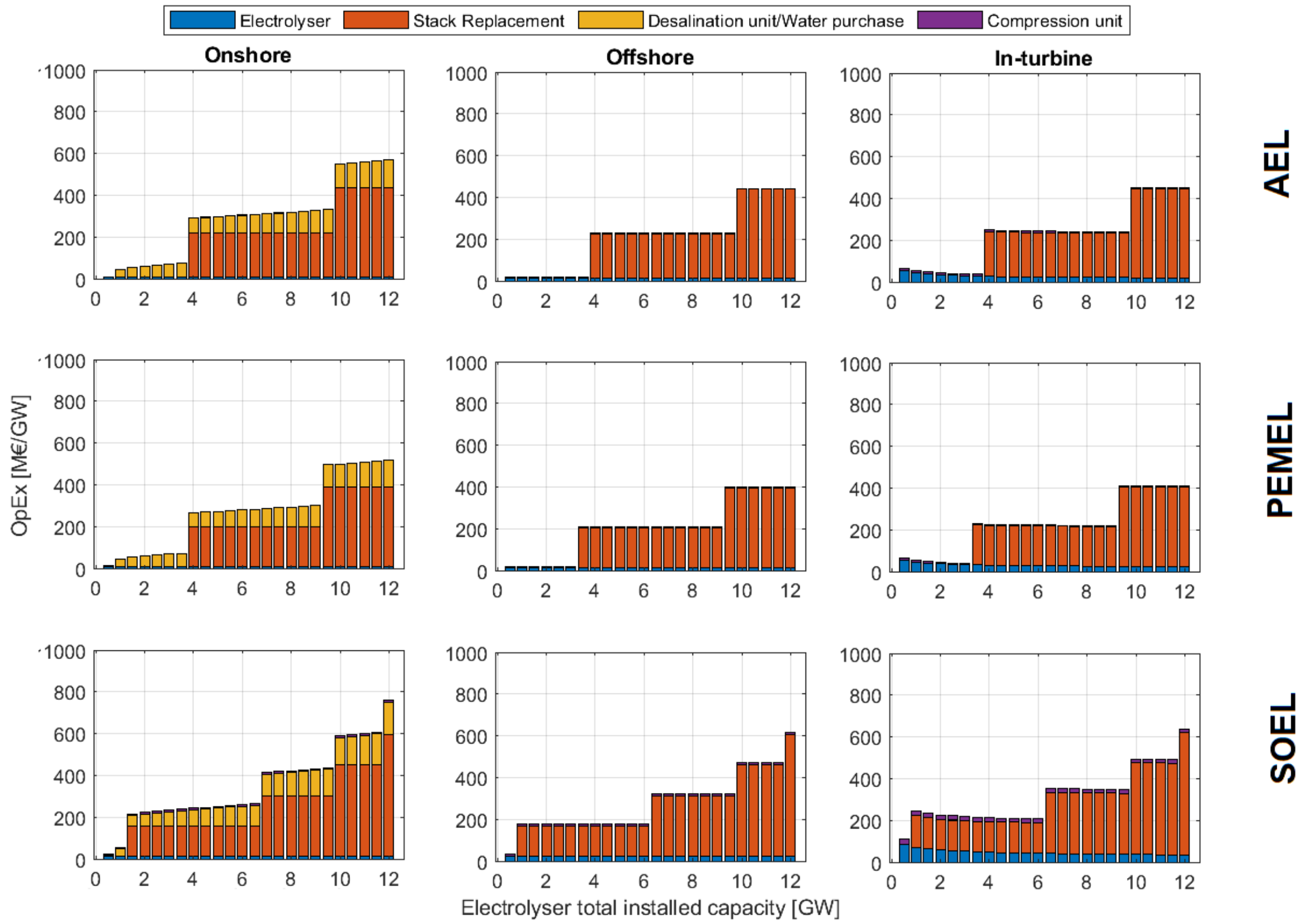

**Figure 6**. OpEx per unit of installed capacity, in the case of electricity-driven operation.

The energy consumption and the energy delivered in form of $H_2$ per unit of installed capacity in **Figure 7**. Both energy consumption and delivered increases by increasing the scale, due to the increasing CFs. AEL and PEMEL show similar performance. SOEL is penalised by the slower cold start-up and by the higher degradation. Moreover, the consumption of the compressor for SOEL is higher than for AEL and PEMEL, because of the higher pressure ratio, therefore less electricity is converted to $H_2$. Due to the assumption for the electricity-driven operation that the priority is to cover the electrical demand onshore, all the losses in the electricity infrastructure upstream of the electrolyser are considered to be allocated in the part of electricity dedicated to the electrolyser. Therefore, the energy input decreases from the in-turbine to the onshore placement, due to the electric losses in the offshore electricity infrastructure. Due to the losses in the electricity infrastructure, in the onshore smallest case (i.e. assumed 500 MW in the model), no energy is consumed/generated by the AEL, and only an irrelevant portion in the PEMEL and SOEL, due to the wider load operational range.

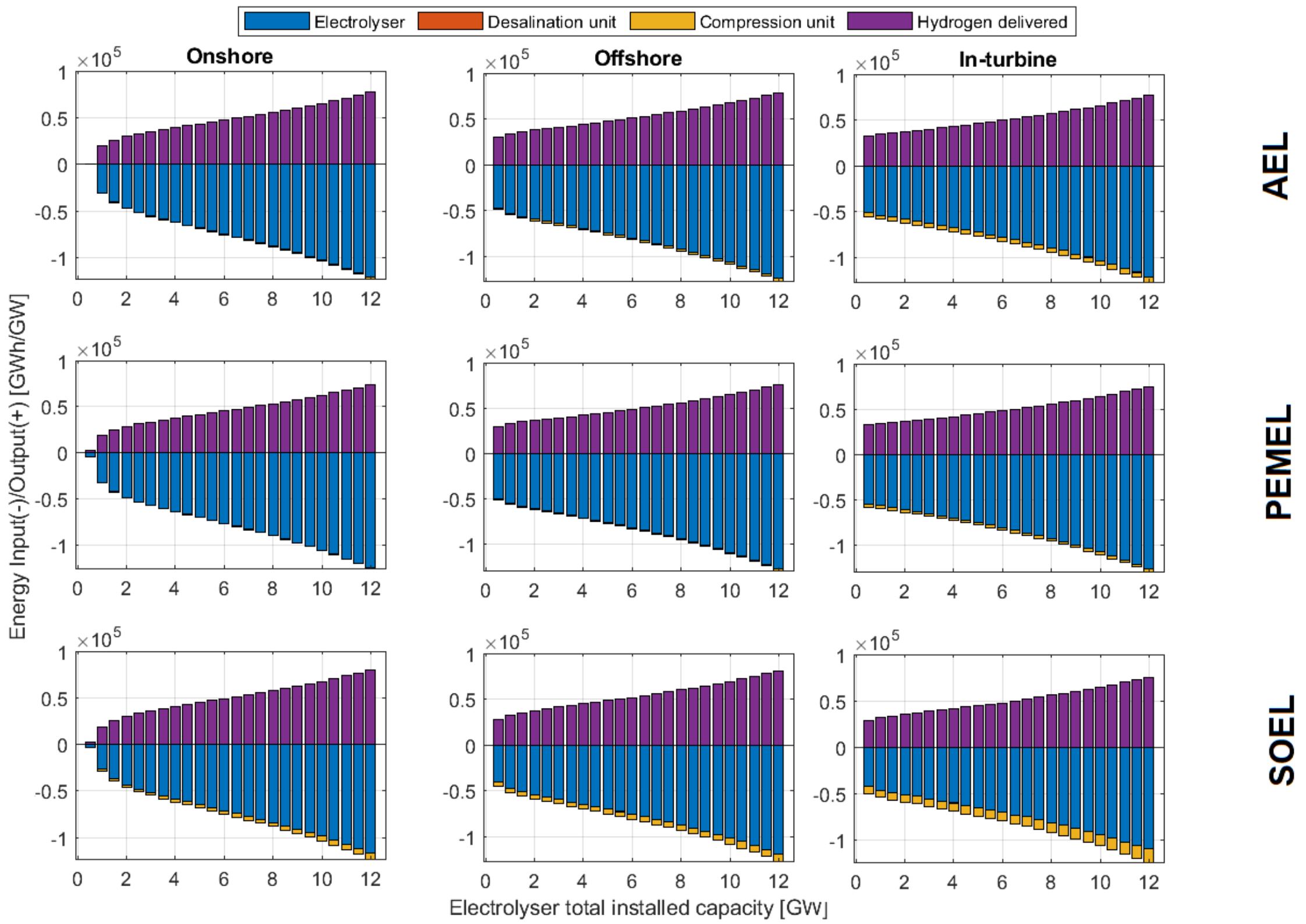

**Figure 7**. Energy consumed in form of electricity and delivered in form of $H_2$, in the case of electricity-driven operation.

The cost of purchasing electricity per unit of installed capacity is presented in **Figure 8**. As in the case of $H_2$-driven operation, this is an order of magnitude higher than the OpEx, and also greater than CapEx, resulting in the main cost driver. The cost for the purchased electricity per unit of installed capacity increases by increasing the installed capacity, due to the increasing CFs. As in the case of $H_2$-driven operation, the cost for the purchased electricity is lower for the in-turbine placement, followed by the offshore and, finally, for the onshore placement. This is due to the cost of the electrical infrastructure upstream the electrolyser allocated in the cost of the electrical energy consumed by the $H_2$ infrastructure, larger for the onshore placement, followed by the offshore and in-turbine placements.

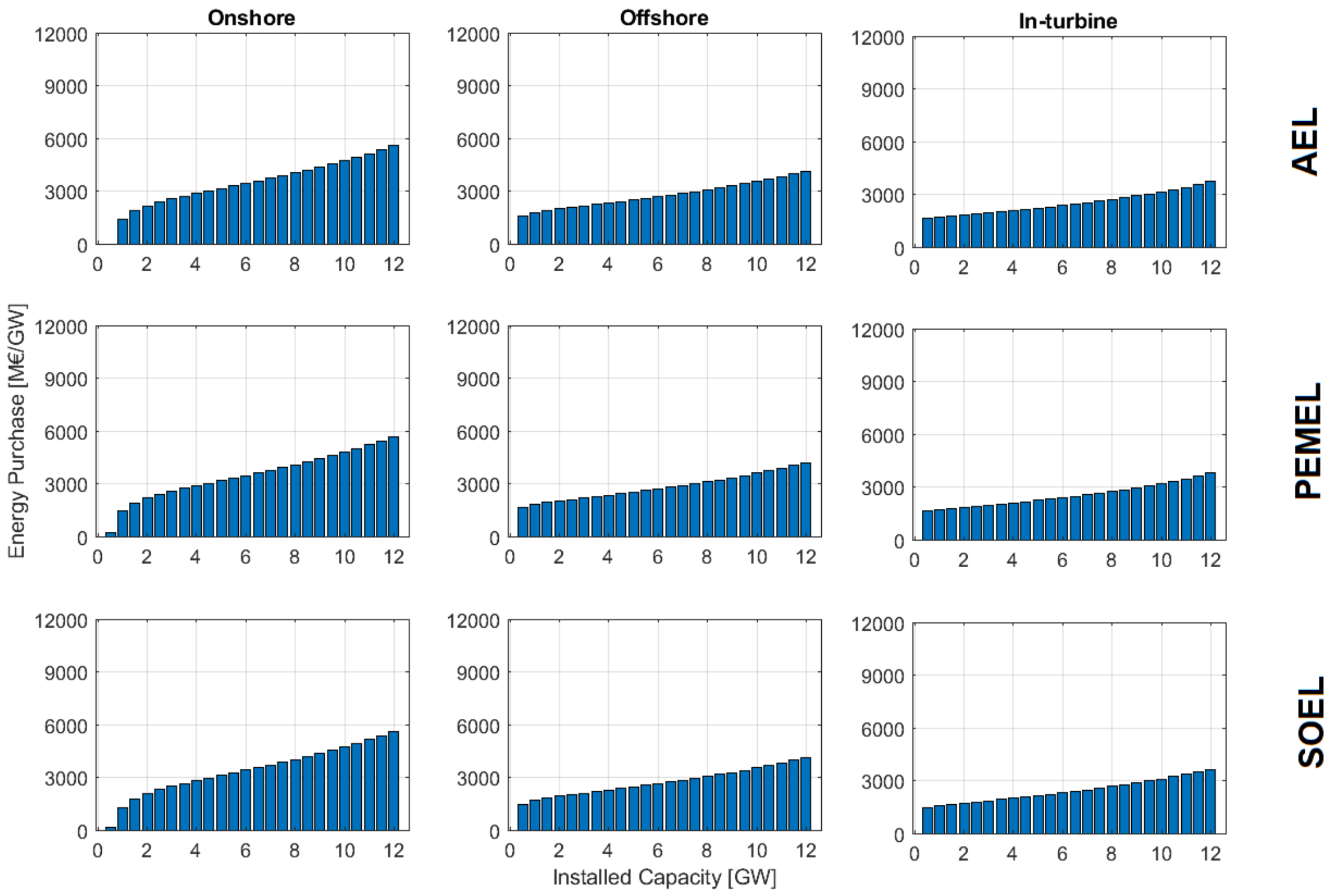

**Figure 8**. Electrical energy purchase per unit of capacity installed, in the case of electricity-driven operation.

The final results of LCOH and LCOE for each electrolyser technology and placement are presented in **Figure 9**. The LCOH is the results of the interrelated effects of CapEx, OpEx, and electrical energy purchase and energy delivered. Lower LCOHs are found for AEL, even though very close to PEMEL. The differences between the alternative technologies are due to the same reasons for the $H_2$-driven operation (see the previous section). Moreover, we found that for central values of the electrolyser total installed capacity (i.e. from 2.5 GW to 6.5 GW) the LCOH of onshore SOEL is lower than the LCOE of offshore SOEL.

LCOE and CF of the power transmission to shore are independent of the electrolyser technology used. Therefore, the discussion regarding the electricity infrastructure in the case of AEL (in the Article) can be extended to PEMEL and SOEL.

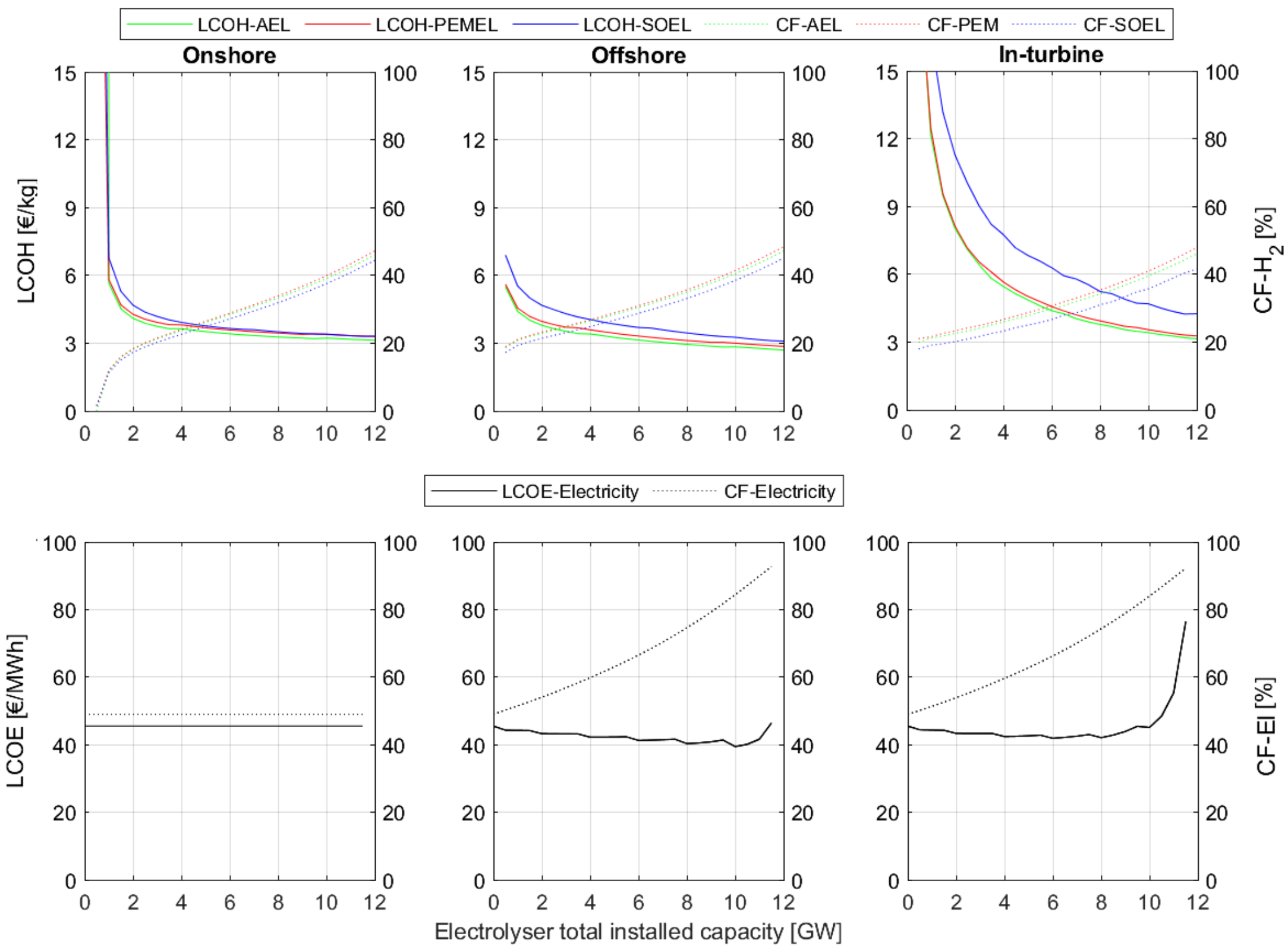

**Figure 9.** LCOH, LCOE and CF, in the case of electricity-driven operation. CF-electricity is the capacity factor of the electricity infrastructure (e.g. HVDC transmission cable, HVDC converters, substations, etc.) from the Hub to shore.

## 2.3 Sensitivity analysis

The effects on the median LCOH of each cost of each component is presented in **Figure 10** and **Figure 11**. The effects of the cost of each technology unit on LCOH already discussed for AEL in the main article are similar to PEMEL and SOEL, with a larger impact of the compression unit for the latter electrolyser type.

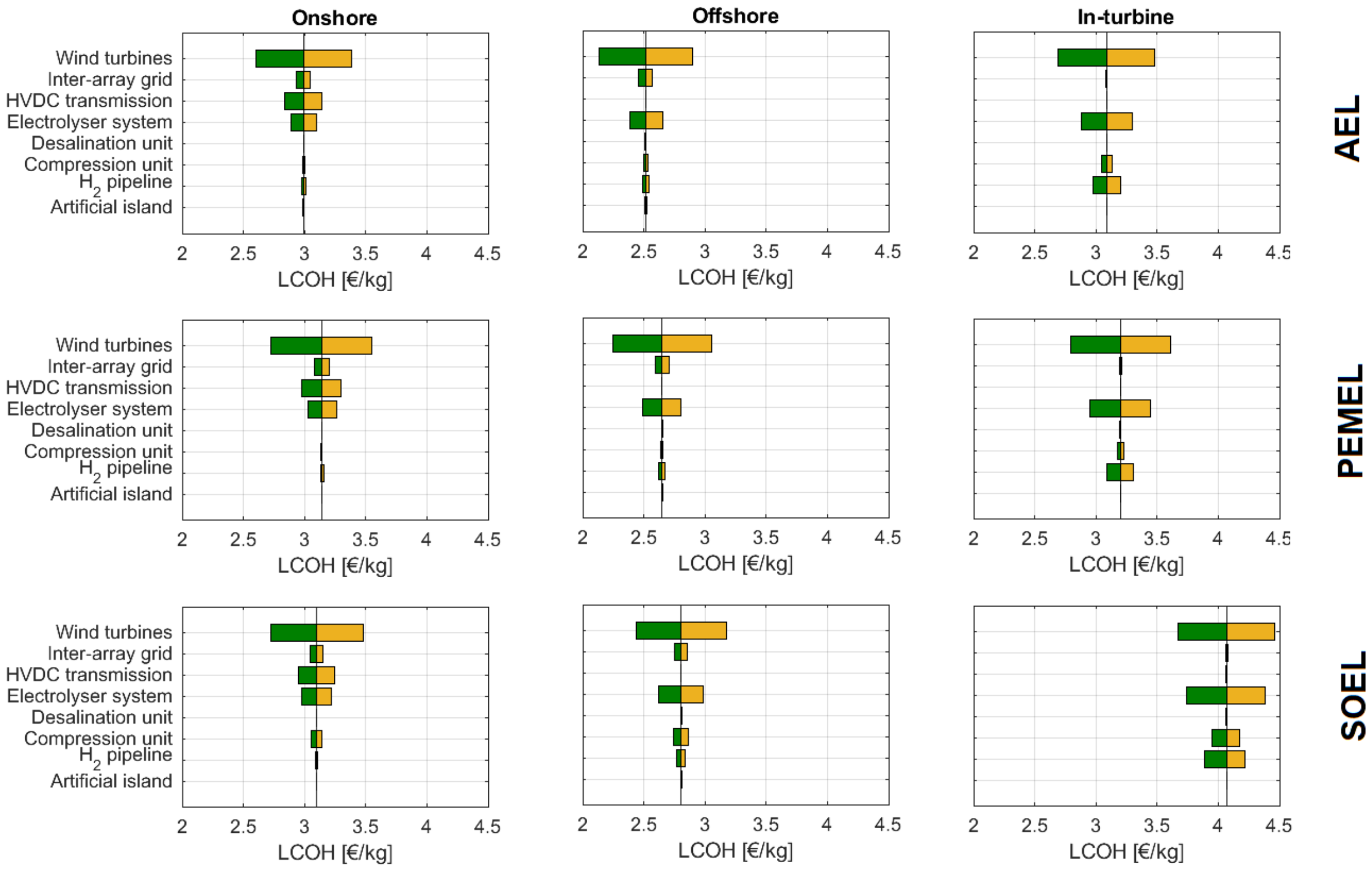

**Figure 10.** Median LCOH obtained by perturbation of +/-25% the CapEx of each component, in the case of $H_2$-driven operation.

In the case of the electricity-driven operation (**Figure 11**), it is important to notice that the median value of LCOH for onshore SOEL is lower than the offshore value of LCOH for offshore SOEL.

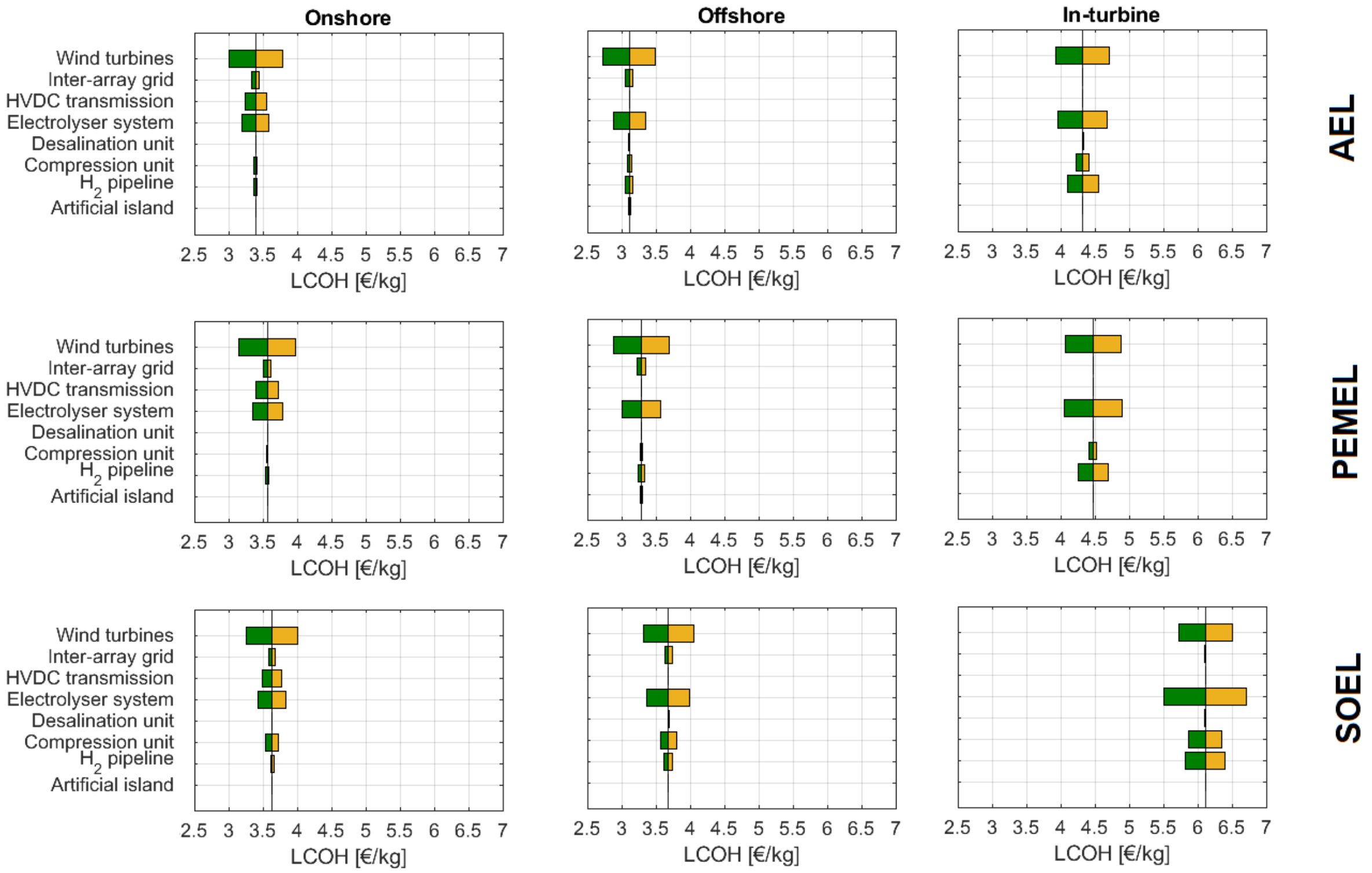

**Figure 11.** Median LCOH obtained by a perturbation of +/-25% the CapEx of each component, in the case of electricity-driven operation.

## 2.4 Impact of the distance of the Hub

The relation between the LCOH and the distance of the Hub from the shore is presented in **Figure 12** and **Figure 13**. The effect of distance on LCOH already discussed for AEL in the main article are similar to PEMEL and SOEL, with higher costs for the in-turbine case of the latter technology, due to the additional compression required.

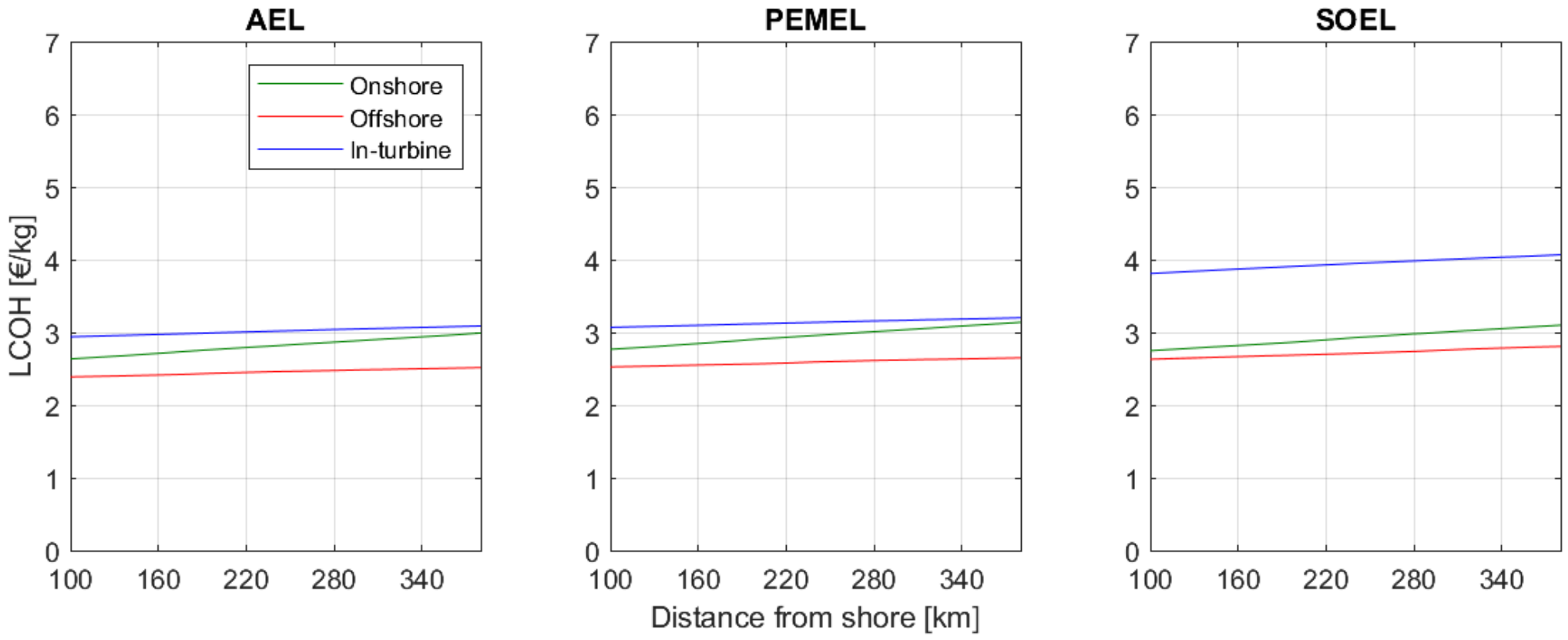

**Figure 12.** Median LCOH by the distance of the Hub to shore, in the case of $H_2$-driven operation.

In the case of electricity-driven operation (**Figure 13**), the proximity to shore makes the onshore solution more competitive with the offshore solution. In the case of SOEL, the onshore placement is more cost-effective than the offshore placement.

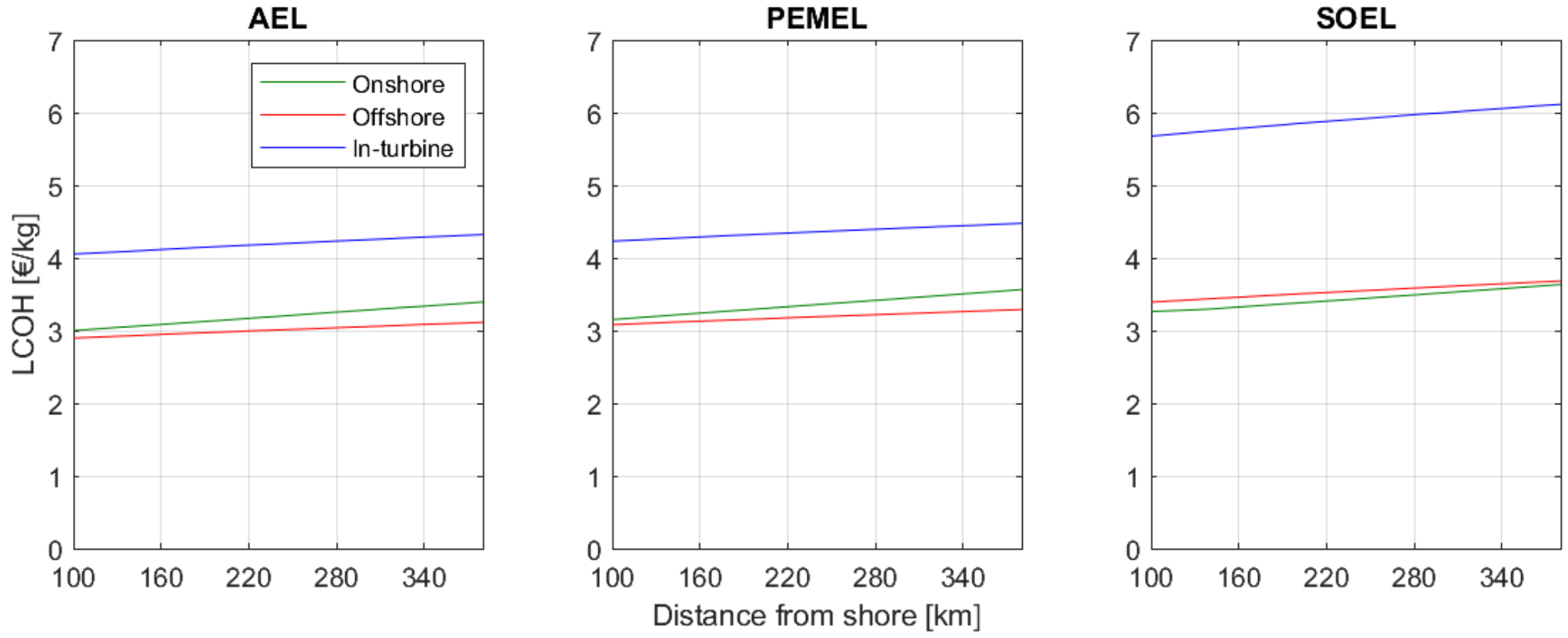

**Figure 13.** Median LCOH by the distance of the Hub to shore, in the case of electricity-driven operation.

## 2.5 Discussion on the footprint.

AEL is the technology with the largest footprint, occupying in the extreme case (i.e. 12 GW), 1.14 $km^2$, in the case of the centralised offshore or offshore placement, or 1,425 $m^2$ in case of the in-turbine solution.

Considering a WT tower of 10 m of diameter [1], the available horizontal area would be 78.5 $m^2$. This would be enough to contain an AEL of 840 kW (without considering the desalination unit). The in-turbine design would be feasible with more compact designs of the electrolyser, otherwise, the electrolyser should be placed outside the WT.

**References**

[1] Gaertner E, Rinker J, Sethuraman L, Anderson B, Zahle F, Barter G. IEA Wind TCP Task 37: Definition of the IEA 15 MW Offshore Reference Wind Turbine. United States: 2020. doi:doi:10.2172/1603478.

[2] E.C.M. Ruijgrok PhD, E.J. van Druten MSc BHBMs. Cost Evaluation of North Sea Offshore Wind Post 2030. Petten, The Netherlands: 2019. doi:112522/19-001.830 112522.

[3] The Danish Energy Agency, Energinet. Technology Data - Generation of Electricity and District heating 2016:414.

[4] Greedy, Lyndon. TENNET, NL OFFSHORE WIND FARM TRANSMISSION SYSTEMS 66 kV Systems for Offshore Wind Farms 2015:35.

[5] U.S. Department of Energy. Assessing HVDC Transmission for Impacts of Non - Dispatchable Generation 2018:1–32.

[6] Tractebel E, Engie, Hinicio. Study on Early Business Cases for H2 in Energy Storage and More Broadly Power To H2 Applications. EU Comm 2017:228.

[7] IEA. The Future of Hydrogen - Seizing today's opportunities. Rep Prep by IEA G20, Japan 2019. doi:10.1787/1e0514c4-en.

[8] Renz M, Schot M Van, Jepma C. North Sea Energy Energy transport and energy carriers 2020.

[9] Electricity Ten Year Statement 2015. Appendix E: Electricity Ten Year Statement 2015:2015. doi:10.1016/B978-0-08-091906-5.00027-6.

[10] Swamy SK, Saraswati N, Warnaar P. North Sea Wind Power Hub ( NSWPH ): Benefit study for ( 1 + 3 ) potential locations of an offshore hub- island. Petten (The Netherlands): 2019. doi:06.37770.

[11] Jepma C, Kok G-J, Renz M, van Schot M, Wouters K. North Sea Energy D3.6 Towards sustainable energy production on the North Sea-Green hydrogen production and CO2 storage: onshore or offshore? As Part of Topsector Energy: TKI Offshore Wind & TKI New Gas: 2018.